 \documentclass[final,5p,times,twocolumn]{elsarticle}

\usepackage{graphicx}
\usepackage{amsmath,amssymb,eqnarray}
\usepackage{mathrsfs}

\journal{European Journal of Mechanics / B Fluids}

\begin{document}

\begin{frontmatter}



\title{Asymmetric vertical transport in weakly forced shallow flows}


\author{L. M. Flores Ram\'irez}
\author{L. P. J. Kamp}
\author{H. J. H. Clercx}
\author{M. Duran-Matute}

\affiliation{organization={Fluids and Flows group and J.M. Burgers Center for Fluid Mechanics, Department of Applied Physics and Science Education},
            addressline={Eindhoven University of Technology}, 
            city={Eindhoven},
            postcode={P.O. Box 513, 5600 MB}, 
            country={The Netherlands}}

\begin{abstract}
In this paper, we report on an investigation of the vertical transport of tracer particles released within a shallow, continuously-forced flow by means of numerical simulations. The investigation is motivated by the shallow flows encountered in many environmental situations and inspired by the laboratory experiments conducted in electromagnetically forced shallow fluid layers.  The flow is confined to a thin fluid layer by stress-free top and no-slip bottom walls. The dynamics and the transport properties of the shallow flow are investigated under various flow conditions characterized by a Reynolds number related to the forcing, $Re_F$, and the aspect ratio of vertical and horizontal length scales $\delta$. The simulated shallow flows exhibit distinctive spatial distributions of vertical velocities: broader, weaker upwellings surrounded by narrower, stronger downwellings. These vertical flows are related to the horizontal flow structures, with updrafts occurring where the horizontal flow is vorticity-dominated, and the downdrafts where it is strain-dominated. The magnitude of the asymmetry in strength and size of the vertical flows and their correlation with horizontal structures depends on the flow conditions and significantly influences the vertical spreading of particles within the fluid volume. Under conditions leading to a large asymmetry, particles within updrafts are transported slowly upwards, while particles within downdrafts rapidly move downwards. In addition, particles are trapped for longer within the updrafts than downdrafts because of their correlation with vorticity-dominated regions. However, when the flow becomes fully three-dimensional and highly unsteady, this transport asymmetry subsides because the updrafts and downdrafts exhibit similar strength and size in such flow conditions.  Consequently, similar amounts of particles are transported upwards and downwards at similar rates.
\end{abstract}




\end{frontmatter}



\section{Introduction}

Shallow flows occur in a wide range of natural environments where the depth of the fluid is much smaller than the horizontal extent \cite{Jirka2001LargeFlows,2004ShallowFlows,VanHeijst2014}.  Some shallow estuaries and lakes have a typical horizontal extent of hundreds of kilometers while the depth is in the order of meters, typically 10-100 m \citep{Clercx2022Quasi-2DLayers}. Due to this vertical confinement, the shallow flow exhibits two-dimensional (2D) characteristics, in particular, the ability to self-organize into coherent vortical structures \cite{Jirka2001LargeFlows,Nikora2007Large-scaleFlows,Stocchino2010}. For example, the flow downstream of a grid of obstacles in a shallow layer exhibits horizontal vortices of size much larger than the water depth \cite{Uijttewaal2003}. These vortices also appear in the far field when two streams, such as rivers, with different velocities converge \cite{Uijttewaal2000EffectsLayers}. However, three-dimensional (3D) effects persist in the form of small-scale turbulent motions, which sometimes destabilize or even destroy the large-scale vortical structures. Importantly, the emergence of the vortices significantly impacts the transport of tracers within the fluid layer because their large scale enables more effective horizontal mixing than the 3D turbulent motions alone \cite{Rummel2005EnhancedTurbulence}, and their capacity to retain high concentrations of tracers \cite{Chen1999LIFLayer}. As an example, in shallow compound channels, vortices produce an intense lateral dispersion of tracer particles from the main channel to the floodplains \cite{Stocchino2011}.

Besides horizontal dispersion, vertical transport plays a critical role in the exchange of, for example, plastics, larvae, or other biogeochemical tracers within environmental shallow flows. In the shallow, coastal ocean, distinct distributions of sinking and buoyant particles in the water column are observed, depending on the turbulent regime, whether wind or wave-driven (each one generating different vertical velocity distributions) \cite{Thoman2021DispersionOcean}. Moreover, recent studies have shown that ignoring the vertical diffusion and advection of microplastics in the transport modeling leads to inaccurate estimations of the horizontal distribution and final destination of plastics in a small bay \cite{Jalon-Rojas2019TechnicalMicroplastics}. These results suggest that the vertical displacement of particles can simultaneously affect the horizontal dispersion. Indeed, vertical motions permit the particles to sample several horizontal sections which differ in their horizontal velocity. For instance, vertical diffusion intensifies the horizontal flux of particles from the interior of an idealized geophysical vortex towards the region outside the vortex \cite{Koshel2015}. It is then clear that vertical dispersion in shallow flows contributes substantially to the overall dispersion, hence the relevance of understanding the relation between the vertical dispersion of tracers and the 3D structure of complex shallow flows. 

In this paper, we study the vertical transport in an idealized, generic shallow flow from a Lagrangian perspective. Despite being idealized, our flow captures generic features of environmental shallow flows that motivate this study: the shallowness, naturally, and more significantly, the presence of large-scale vortices. Our investigation is further inspired by the laboratory experiments conducted in electromagnetically forced shallow fluid layers.  These experiments in combination with 3D numerical simulations have shown that these flows have a complex unsteady 3D structure in spite of their shallowness \cite{Cieslik2009,Cieslik2010,Akkermans2012ArraysDispersion}. To understand these flows, there has been extensive research on shallow monopolar and dipolar vortices, which can be considered as their building blocks \cite{Satijn2001Three-dimensionalLayers,Sous2004TurbulentLayer,Akkermans2008,Duran-Matute2010ScalingFlows,Duran-Matute2011,Kamp2012,Albagnac2014ADipole}. 

In the case of a monopolar vortex, the 3D structure arises from the vertical shear of its horizontal velocity field. This shear, caused by the no-slip bottom, creates an imbalance between the centrifugal force and the horizontal pressure gradient within the primary vortex. This imbalance drives secondary motions with vertical and radial velocity components \cite{Satijn2001Three-dimensionalLayers,Duran-Matute2010ScalingFlows}.  Specifically, there is an upward motion (updraft or upwelling) inside the vortex core with radial inflow close to the bottom wall and radial outflow near the free surface. Due to mass conservation, a downward motion (downdraft or downwelling) forms at the vortex periphery with radial inflow close to the free surface and a radial outflow close to the bottom \cite{Bodewadt1940DieGrunde,Satijn2001Three-dimensionalLayers,Kamp2012}. The magnitude of the radial and vertical velocities scale with the parameters $Re\delta^2$ and $Re\delta^3$, respectively, where $Re=UR/\nu$ is the Reynolds number, $\delta=H/R$ is the aspect ratio, with $U$ a typical velocity scale, $R$ the vortex radius, $H$ the fluid layer depth, and $\nu$ the kinematic viscosity of the fluid \cite{Duran-Matute2010ScalingFlows}. The importance of these secondary motions is due to their capacity to modify the primary motion if they are strong enough. For $Re\delta^2 \lesssim 6$ (or $Re \delta^3\lesssim 1$), the shallow vortex is dominated by vertical viscous diffusion, and secondary flows are negligible. However, if $Re\delta^2\gtrsim 6$ (or $Re \delta^3\gtrsim  1$), the primary vortex is affected by the secondary motions \cite{Duran-Matute2010ScalingFlows}. 

Although more complex, dipolar vortices show similar behavior \cite{Duran-Matute2010DynamicsVortices}. In particular, upward motion forms in the two primary vortex cores surrounded by negative vertical velocity similar to the
secondary circulation of the monopolar vortex. In addition, when the value of $Re\delta^2$ is large enough, a frontal circulation is observed, that is a roll-like structure with upward motion at the front of the dipole followed by a weak downward motion \cite{Sous2004TurbulentLayer,Akkermans2008,Duran-Matute2010DynamicsVortices,Albagnac2014ADipole}. 

For decaying, multiple-vortex shallow  flows, where monopolar and dipolar vortices are continuously interacting, vertical motions get organized in a complex, asymmetric distribution inside the layer \cite{Cieslik2009,Cieslik2010}. This asymmetry is evident in both the magnitude of the vertical velocity and the area occupied by the vertical movements, and it is correlated with the horizontal velocity field. In particular, intense downwellings are observed in thin, elongated patches within strain-dominated regions of the horizontal flow, whereas weak upwellings occupy more extended patches that are correlated with rotation-dominated regions of the horizontal flow \cite{Cieslik2009,Cieslik2010,Kamp2012}. Similar spatial distribution of vertical motions has been observed in tidal shallow wakes \cite{Branson2019Three-dimensionalityWakes} and in an array of shallow, periodically forced vortices \cite{Akkermans2012ArraysDispersion}. 

Overall, an intricate spatial distribution of the upwellings and downwellings characterizes these complex shallow flows. The form in which these vertical motions are distributed determines how materials are vertically transported within the layer. For instance, in shallow dipolar vortices, the upward motions located at the vortex cores and at the frontal circulation allow the dipole to carry material from the bottom and put it into suspension in the fluid layer, similarly to a mechanical carpet sweeper removing dust \cite{Albagnac2014ADipole}. The goal of this work is to determine the relation between the vertical spreading of passive tracers and the distribution of vertical structures in shallow flows by means of 3D numerical simulations. To achieve this, we analyze the spatial distribution of updrafts and downdrafts in our simulated flows, explore how this distribution changes with the control parameters of the flow, and assess the impact of such changes on the relation between updrafts and downdrafts with the horizontal structures of the flow.  With this information in hand, we explore the dispersive characteristics of updrafts and downdrafts as they advect particles throughout the shallow layer.

The paper is organized as follows. In Section \ref{sec:num_setup}, the problem is formulated in detail and the methods are described. Section \ref{sec:charac_euler} is devoted to the characterization of the velocity field of the flow from an Eulerian perspective. Section \ref{sec:vert_transp} focuses on the vertical transport of particles induced by such flows. Finally, Section \ref{sec:discu_conclu} discusses the findings of this study and presents the main conclusions.

\section{Statement of the problem and numerical set-up of the simulations}\label{sec:num_setup}

We consider a shallow flow with depth $H$ that is vertically confined by rigid boundaries (a stress-free top and a no-slip bottom). The depth $H$ is much smaller than the horizontal extent of the domain $L$. The flow is continuously forced with the forcing being reminiscent of the electromagnetic forcing commonly used in shallow electrolytic solutions \cite{Sommeria1986ExperimentalBox,Tabeling1987,Paret1997ExperimentalCascade,Clercx2003Quasi-two-dimensionalDepth,Kelley2011,Tithof2018}. We study the flow in a statistically steady state, which consists of multiple vortices interacting with each other like in the aforementioned experiments. The forcing has a typical horizontal length scale $L_f$ with $L_f>H$. To study the Lagrangian transport properties, we release passive particles at mid-depth and perform a statistical analysis using the information of the particles position and velocities. Further details on the implementation are given below. 

\subsection{Governing equations} 

The flow is governed by the Navier-Stokes equation:
 
\begin{equation} 
\frac{\partial\mathbf{u}}{\partial t}+(\mathbf{u}\cdot\boldsymbol{\nabla})\mathbf{u}=-\frac{1}{\rho}\boldsymbol{\nabla} p+\nu\nabla^2\mathbf{u}+\dfrac{\mathbf{f}}{\rho}, \label{eq:ns3d}
\end{equation}

\noindent and the continuity equation for an incompressible fluid:

\begin{equation}
\boldsymbol{\nabla}\cdot\mathbf{u}=0, \label{eq:cont}
\end{equation}

\noindent where $\mathbf{u}=(u,v,w)$ is the flow velocity, $t$ is the time, $\rho$ is the fluid density, $p$ is the pressure, $\nu$ is the kinematic viscosity, and $\mathbf{f}$ is the external body force per unit volume. We adopt a Cartesian reference frame $(x,y,z)$, with $x$ and $y$ the horizontal directions and $z$ the vertical one. 

To nondimensionalize Eqs.~\eqref{eq:ns3d} and \eqref{eq:cont}, we define the following non-dimensional variables (denoted by primes): 
\[\mathbf{u}'=\mathbf{u}/\mathcal{U},\quad t'=\mathcal{U} t/L_f, \quad (x',y',z')=(x,y,z)/L_f,\quad\] 
\[p'=p/(\rho \mathcal{U}^2),\quad \mathbf{f}'=\mathbf{f}/\mathcal{F},\]
where $\mathcal{U}$ is a typical velocity scale and $\mathcal{F}$ is the forcing magnitude. Here $\mathcal{U}$ is set equal to $\sqrt{\mathcal{F}L_f/\rho}$, which means that inertia forces are considered of the same order as the imposed body force. Substituting these variables into Eqs.~\eqref{eq:ns3d} and \eqref{eq:cont}, we obtain the non-dimensional governing equations, which after dropping the primes, read

\begin{equation}
    \frac{\partial\mathbf{u}}{\partial t}+(\mathbf{u}\cdot\boldsymbol{\nabla})\mathbf{u}=-\boldsymbol{\nabla} p+\frac{1}{Re_F}\nabla^2\mathbf{u}+\mathbf{f}, \label{eq:ns3d_ndim}
\end{equation}

\begin{equation}
    \boldsymbol{\nabla}\cdot\mathbf{u}=0, \label{eq:cont_ndim}
\end{equation}

\noindent where $Re_F= (L_f/\nu)\sqrt{\mathcal{F} L_f/\rho}$ is the Reynolds number based on the forcing magnitude as also defined by, for example, \citet{Alexakis2015RotatingFlow}. Once the equations are solved, a Reynolds number $Re$ based on an actual horizontal  and vertical velocity scale of the flow can be computed. Here, we define the Reynolds number for the horizontal velocities as

\begin{equation}
    Re=\dfrac{U_{\mathrm{rms}}L_f}{\nu}=\dfrac{U_{\mathrm{rms}}}{\sqrt{\mathcal{F}L_f/\rho}}Re_F, \label{eq:ReF2Re}
\end{equation}

\noindent with $U_{\mathrm{rms}}$ the root-mean-square (r.m.s) velocity at the top, stress-free boundary, and for the vertical velocities as 
\begin{equation}
    Re_w=\dfrac{W_{\mathrm{rms}}H}{\nu}=\dfrac{W_{\mathrm{rms}}}{\sqrt{\mathcal{F}L_f/\rho}}Re_F\delta,
\end{equation}
with $W_\mathrm{rms}$ the r.m.s vertical velocity at mid-depth, and $\delta=H/L_f$.

The dimensionless body force  $\mathbf{f}$ is defined as

\begin{equation}
    \mathbf{f}=(f_x,f_y,f_z)=\dfrac{1}{2\pi^2} \left( \dfrac{\partial q}{\partial y},-\dfrac{\partial q}{\partial x},0 \right)
\end{equation}

\noindent with

\begin{equation}
    q=\sin(\pi x)\sin(\pi y),
\end{equation}

\noindent which implies that $\boldsymbol{\nabla}\cdot\mathbf{f}=0$. This is a steady forcing with fixed amplitude active at a single scale. This scale is equal to unity in the non-dimensional form, since we have used the forcing scale $L_f$ as the typical length scale. Furthermore, note that the horizontal components of the forcing do not have a vertical dependence and that the flow is not forced in the vertical direction. These are important differences with respect to electromagnetically forced shallow flows 
\cite{Tabeling1987,Kelley2011,Tithof2018}. The total number of vortices being initially forced is determined by the relative horizontal size of the domain $L/L_f$ so that $L$ must be a integer multiple of $L_f$, specifically an even integer (due to periodic boundary conditions). 

\subsection{Numerical set-up} 

Numerical simulations are performed by solving the dimensionless governing Eqs.~\-\eqref{eq:ns3d_ndim} and \eqref{eq:cont_ndim} with a finite element code \cite{COMSOL}. The computational domain is a rectangular box of dimensions $4\times4\times \delta$. The flow is vertically confined by rigid boundaries. A stress-free boundary condition is applied to the upper boundary, and free-surface deformations are not taken into account, while a no-slip boundary condition is applied at the bottom. The lateral boundaries are periodic. We explore the parameter space by conducting the simulations for three values of the aspect ratio, $\delta=0.1,\, 0.3$ and $0.5$, and the different values of the Reynolds number $Re_F$ shown in Table \ref{table:param}. After reaching a statistically steady state, the corresponding flow Reynolds number $Re$ is computed using Eq.~\eqref{eq:ReF2Re}.  

The domain was discretized using a mesh composed of triangular-prism elements. This type of mesh is convenient to discretize geometries with a small aspect ratio $\delta$. An unstructured mesh, formed by triangular elements, is used to discretize the horizontal directions. The vertical direction is discretized with a structured mesh made of rectangular elements, with finer elements placed close to the top and bottom boundaries. The number of mesh elements in the horizontal and vertical directions changes depending on the value of $Re_F$ and $\delta$ to achieve accurate results with the least computational cost. Typically, the total resolution ranged from 40,000 to 700,000 elements, with 20 to 30 elements in the vertical direction. To confirm the adequacy of the grid, we verified the zero divergence condition, Eq.~\eqref{eq:cont_ndim}, which is susceptible to insufficient resolution. Specifically, we computed the mean absolute value of the divergence to evaluate the simulation accuracy. For most of our cases, the divergence was approximately 1-2\% of the r.m.s vorticity, which was considered sufficient. Besides, the convergence of the results was tested for different parameter values by running the simulations with additional mesh refinement, after which the result appears indistinguishable. The time stepping uses variable-order, variable-step-size backward differentiation formulas, with the order and the step size determined by the numerical code based on the simulation requirements \cite{COMSOL}. When the flow is statistically steady, the time step becomes approximately constant. 

\begin{table*}
\centering
\caption{Values of the aspect ratio $\delta$, the Reynolds number based on the forcing magnitude, $Re_F$, used in the numerical simulations (both control parameters), and the corresponding value of the obtained flow Reynolds number $Re$ (global response of the system; $Re(Re_F,\delta)$).}
\label{table:param}
\begin{tabular}{c|c|c}
\hline
$\delta$ & $Re_F$ & $Re$ \\
\hline
0.1 & 475, 600, 1000, 1650, 2825, 4000, 6000 & 111, 167, 343, 642, 1213, 1736, 2574  \\
\hline
0.3 & 70, 115, 220, 325, 430, 745, 1060 & 14, 34, 82, 129, 185, 326, 470   \\
\hline
0.5 & 35, 70, 90, 130, 175, 290, 400 & 6, 20, 28, 47, 67, 121, 170 \\
\hline
\end{tabular}
\end{table*}

\subsection{Lagrangian particle tracking} \label{sec:LagPartTrack}

The simulated flows were used to obtain the trajectories of passively advected particles that were deployed after the flow reached a statistically steady state. The onset of such state was determined by inspection of the time derivative of the total kinetic energy of the flow. After a transient phase, this quantity  fluctuates around zero over time. The steady state is considered achieved when the magnitude of the fluctuations is consistently maintained below $10^{-1}$ (with the kinetic energy being of order one). The particles were initially located at mid-depth (i.e. $z/\delta=0.5$) while they were randomly distributed in the horizontal. The initial vertical position of the particles was chosen to be mid-depth because the vertical velocities are stronger there, translating to a large vertical particle dispersion. In addition, this choice minimizes the effect of the top and bottom boundary on particle trajectories at early times.

The position $\mathbf{x}^p(t)$ of each particle $p$ is obtained by integrating the equations of motion:

\begin{equation}
    \dfrac{\mathrm{d}\mathbf{x}^p}{\mathrm{d}t}=\mathbf{u}(\mathbf{x}^p,t) \label{eq:lagx}
\end{equation}

\noindent subject to the initial condition $\mathbf{x}^p_0=\mathbf{x}^p(0)$. The particle velocities, $\mathbf{u}^p(t)=\mathbf{u}(\mathbf{x}^p,t)$, are obtained by interpolating the velocity field at the particle position using linear interpolation. The tracking is performed offline, i.e., the flow fields are first simulated and stored and then the particle trajectories are computed using the output velocity fields. The time integration of Eq.~\eqref{eq:lagx} is performed using a fourth-order Runge-Kutta method. To obtain robust statistics and multiple realizations, we deploy $N=50,000$ particles at five different times. The velocity fields at these times are uncorrelated with each other.  In this way, the particles sample a different initial velocity field in each realization, and a total of $N_p=250,000$ particle trajectories are obtained to perform statistical analysis. To ensure we have enough particles to obtain reliable statistics, the relative error in performing the ensemble averaging is calculated. To do so, we compute the difference between the $W_\mathrm{rms}$ value obtained from sub-ensembles of different sizes $M_p$ (from 2 to 62,500 particles) and from the total ensemble of size $N_p$. This error goes towards zero as $M_p^{-1/2}$, consistent with the central limit theorem \cite{Elghobashi1992DirectTurbulence}. Hence, our chosen number of particles yields an acceptable error of $0.2\%$.

\section{Characterization of the Eulerian velocity field}\label{sec:charac_euler}

In this section, as a first step, we characterize the Eulerian velocity. The aim is two-fold: (i) to examine the response of the flow to variations of the governing parameters and (ii) help in the interpretation of the results on the Lagrangian transport. 

\subsection{Viscous and inertial flow regimes}\label{subsec:flow_regimes}

We determine the scaling of the magnitude of the flow velocities in our simulated flows as a function of the parameters $Re_F$ and $\delta$. To do this, we conduct a dimensional analysis of Eq.~\eqref{eq:ns3d} in the limit $\delta\ll 1$, focusing solely on the potential balances of forces within the flow in the horizontal direction. A similar analysis has been previously performed for a dipolar flow structure driven electromagnetically \cite{Duran-Matute2011}.

From this analysis, two different scaling regimes are obtained (see the Appendix for details). In the first regime, $Re\sim Re_F^2 \delta^2/\pi^2$, and viscous diffusion from bottom friction dominates over inertia. In the second one, $Re\sim Re_F$, and in the bulk inertia is of the same order as the forcing.  Moreover, the balance between inertia and viscous forces near the no-slip bottom wall results in a typical vertical scale $h/H\sim (Re_F^{1/2}\delta)^{-1}$, which can be considered as a thickness of the boundary layer at the bottom. The transition between the two regimes occurs when $Re_F^2 \delta^2/\pi^2 \sim Re_F$, that is, when $Re_F\delta^2\sim \pi^2 \approx 10$. 

\begin{figure}[t]
    \centering
    \includegraphics[scale=0.5]{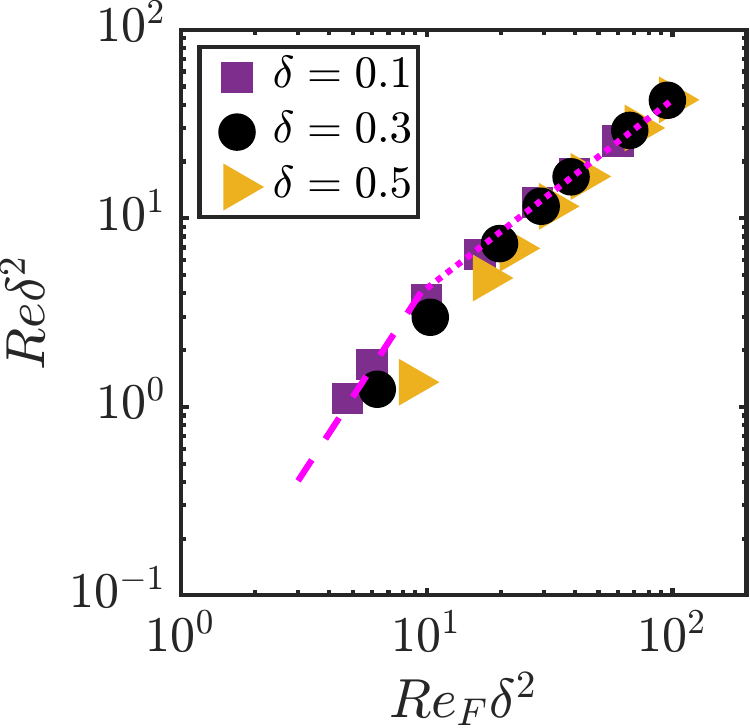}
    \caption{Measured values of $Re\delta^2$ as a function of the parameter $Re_F\delta^2$. The dashed line represents the scaling $Re\sim Re_F^2\delta^2$ (viscous regime) and the dotted line represents the scaling $Re\sim Re_F$ (inertial regime).}
    \label{fig:ReDelta2_vs_ReFDelta2}
\end{figure}

To observe these regimes in our simulations, the  values of $Re \delta^2$ are plotted, in Fig.~\ref{fig:ReDelta2_vs_ReFDelta2}, as a function of $Re_F \delta^2$. The data for different values of $\delta$ collapse onto a single curve. Furthermore, the trends agree with the theoretical prediction of the viscous and inertial regimes, with most of the simulations falling in the inertial regime. The transition at $Re_F\delta^2\approx 10$ is clearly observed for the simulations with $\delta=0.1$, while for the case $\delta=0.5$ it seems to occur for a slightly higher value of $Re_F\delta^2$. This shift can be explained by the additional horizontal viscous effects that are not negligible when the condition $\delta\ll 1$ does not hold. 

\begin{figure*}
    \centering
    \includegraphics[scale=0.5]{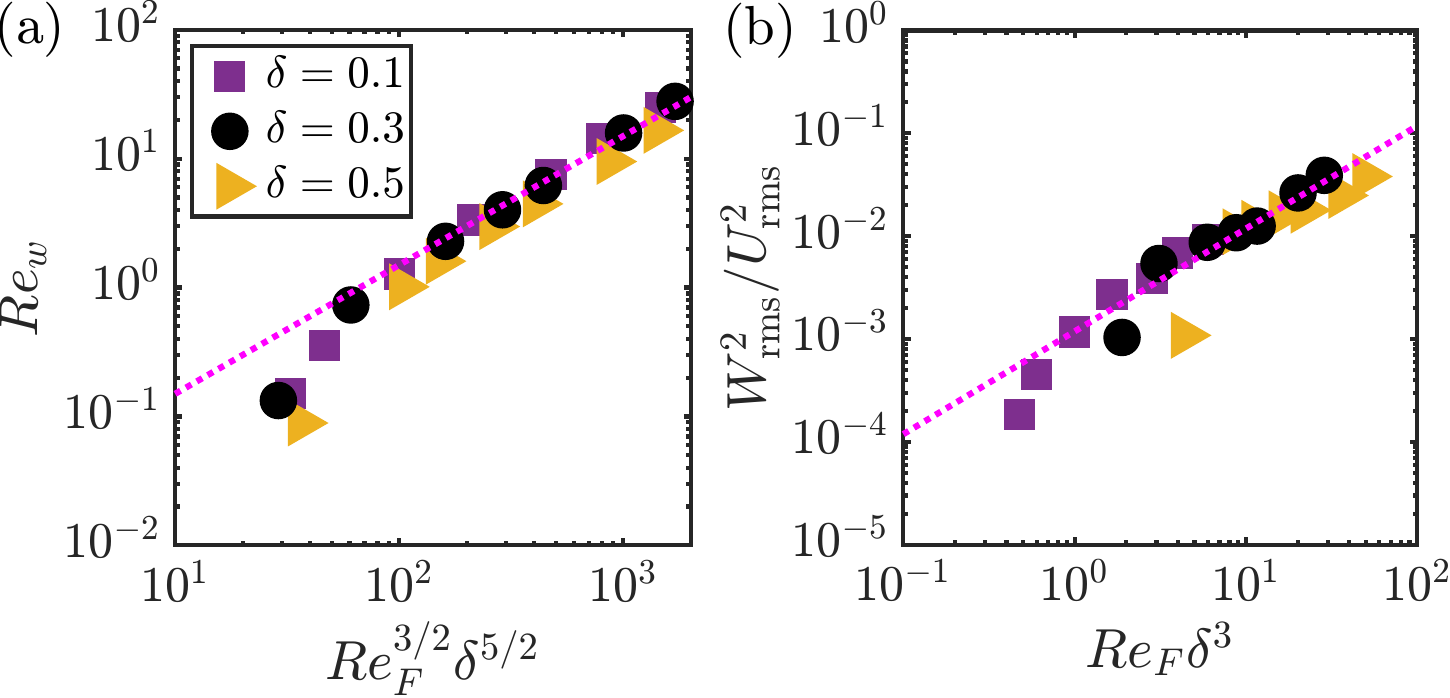}
    \caption{(a) Vertical velocity represented by $Re_w$ as a function of the control parameter $Re_F^{3/2}\delta^{5/2}$. (b) Ratio of kinetic energies $W_\mathrm{rms}^2/U_\mathrm{rms}^2$ as a function of the parameter $Re_F\delta^3$. The dotted lines in both panels represent linear scalings.}
    \label{fig:ReW_vs_ReF1.5Delta2.5}
\end{figure*}

For the vertical velocity, the regimes are not deduced based on  dimensional analysis because simplified balances of forces are not fully realized due to the presence of the no-slip bottom. Instead, regimes are identified based on the obtained values of $Re_w$ as a function of both $Re_F$ and $\delta$.  The values of $Re_w$ exhibit a collapse when plotted as a function of  $Re_F^{3/2}\delta^{5/2}$ (Fig.~\ref{fig:ReW_vs_ReF1.5Delta2.5}a), with the deeper case ($\delta=0.5$) showing a slight offset. For $Re_w > 1$, $Re_w \sim Re_F^{3/2}\delta^{5/2}$. This relation is  equivalent to $W_\mathrm{rms}^2/U_\mathrm{rms}^2 \sim Re_F^3\delta^3/Re^2$. These simulations lie in the inertial regime of the horizontal flow. Since $Re_F \sim Re$ in this regime, it follows that $W_\mathrm{rms}^2/U_\mathrm{rms}^2 \sim Re\delta^3 \sim Re_F\delta^3$. This implies that the ratio of the horizontal and vertical kinetic energies varies linearly with the parameter $Re_F\delta^3$ (as Fig.~\ref{fig:ReW_vs_ReF1.5Delta2.5}b shows). This parameter has also been shown to be the governing parameter for the vertical velocities in shallow monopolar and dipolar vortices \cite{Duran-Matute2010ScalingFlows,Duran-Matute2010DynamicsVortices}. 

\subsection{Three-dimensional structure of the flow}

The vertical structure of the velocity field of the flow is characterized in this section. For this, we compute the vertical profiles of the first three statistical moments (i.e. the mean, the standard deviation and the skewness) of the velocity components. We define these quantities, with the function $g(x,y,z,t)$ representing one of the velocity components, as follows. The mean is given by
\begin{equation}
    \langle g \rangle (z)= \dfrac{1}{T}\int_{0}^T \dfrac{1}{L^2}\int_0^L \int_0^L g(x,y,z,t) \,\mathrm{d}x\,\mathrm{d}y\,\mathrm{d}t, \label{eq:average}
\end{equation}
the standard deviation by
\begin{equation}
    \sigma_g (z) = \langle \left( g - \langle g \rangle \right)^2 \rangle^{1/2},
\end{equation}
and the skewness by 
\begin{equation}
        S_g(z)=\dfrac{\langle  \left( g - \langle g \rangle \right)^3 \rangle}{\sigma_g^3}.
\end{equation}
Note that the averaging is performed over the horizontal directions and over time, starting when the flow is statistically steady ($t=0$) up to the simulation end time $T$. 

Besides, an alternative mean, with the averaging conducted solely over time, is introduced:

\begin{equation}
    \langle g \rangle_t (x,y,z)= \dfrac{1}{T}\int_{0}^T  g(x,y,z,t) \mathrm{d}t. \label{eq:average_time}
\end{equation}
This mean will be used to analyze the statistics of the temporal fluctuations of the velocity field. The fluctuations of the function $g$, which represents one of the velocity components, are defined as $g - \langle g \rangle _t$. Thus, the standard deviation of such fluctuations is given by

\begin{equation}
    \sigma^f_g (z) = \langle \left( g - \langle g \rangle _t \right)^2 \rangle^{1/2}. \label{eq:std_temp}
\end{equation}

\subsubsection{Horizontal velocities}

The mean values of the two horizontal velocity components $\langle u \rangle$  and $\langle v \rangle$ are  numerically zero at every depth because there is no mean flow in the horizontal. Likewise, their skewness $S_u$ and $S_v$ are small (order $10^{-2}$), which indicates that the horizontal velocities are symmetrically distributed around their mean. In other words, there is no preferential direction for the horizontal velocity components.

\begin{figure*}[t]
    \centering
    \includegraphics[scale=0.33]{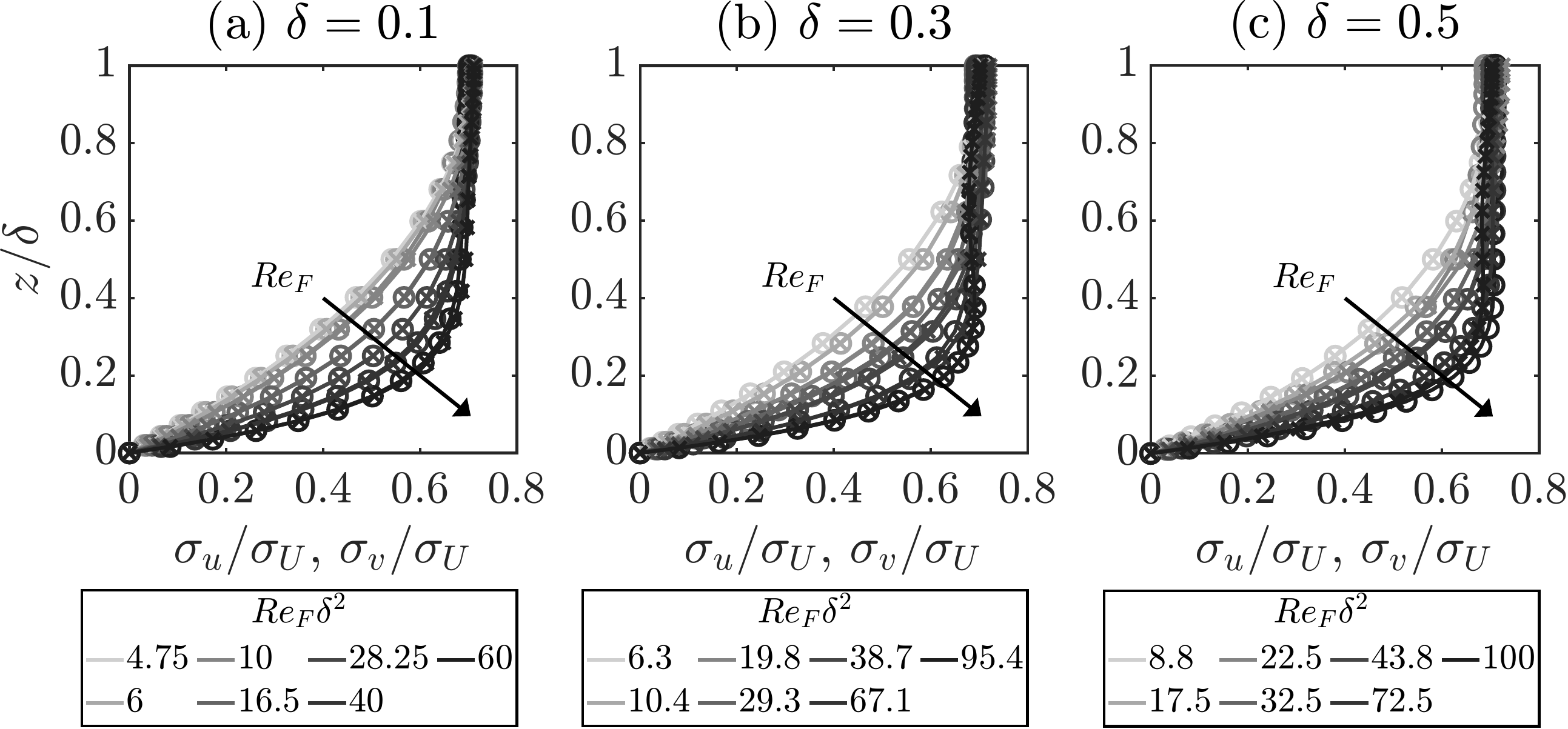}
    \caption{Vertical profiles of the standard deviation of the horizontal velocity components $\sigma_u$ (circles) and $\sigma_v$ (crosses) normalized by the total standard deviation of the horizontal flow at the surface, $\sigma_{U}$. The subfigures show the profiles for the different $\delta$-values as indicated in their title. The different curves represent simulations with different $Re_F$ values with the arrow indicating increasing $Re_F$ values.}
    \label{fig:zprofiles_horizontal}
\end{figure*}

 Fig.~\ref{fig:zprofiles_horizontal} shows the vertical profiles of the standard deviation $\sigma_u=\langle u^2\rangle^{1/2}$ and $\sigma_v=\langle v^2\rangle^{1/2}$ for different aspect ratios $\delta$. The values of $\sigma_u$ and $\sigma_v$ are normalized by the total standard deviation of the horizontal flow at the surface, $\sigma_{U}=\sqrt{[\sigma_u(\delta)]^2+[\sigma_v(\delta)]^2}$. Since the mean value of the velocity is zero, the standard deviation of the components is equivalent to their r.m.s value $u_\mathrm{rms}$ and $v_\mathrm{rms}$, and similarly, the value of $\sigma_{U}$ is equivalent to the r.m.s velocity $U_{\mathrm{rms}}$. For a fixed $\delta$ value, the shape of the velocity profile changes as the value of $Re_F$ (or, by extension, $Re$) changes. In fact, the velocity profiles collapse for similar values of $Re_F\delta^2$. In the viscous regime ($Re_F\delta^2\lesssim10$), the flow has a Poiseuille-like profile due to the dominance of viscous forces. In the inertial regime ($Re_F\delta^2\gtrsim10$), the flow tends towards being composed of an inviscid interior and a boundary layer close to the bottom  with a thickness of order $(Re_F^{1/2}\delta)^{-1}$ (see Appendix \ref{appendix:dimensional}). Finally, note that the profiles of $\sigma_u$ and $\sigma_v$ are nearly identical, thus almost indistinguishable in Fig.~\ref{fig:zprofiles_horizontal}, meaning that the horizontal flow is indeed statistically isotropic in the horizontal.

\begin{figure}[b!]
    \centering
    \includegraphics[scale=0.5]{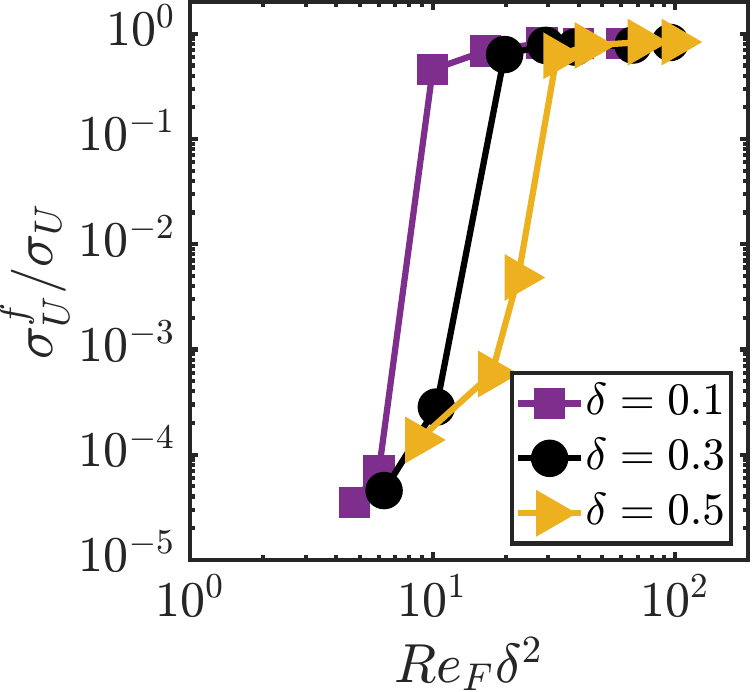}
    \caption{Total standard deviation of the horizontal flow fluctuations at the surface, $\sigma_U^f$, normalized by $\sigma_U$, against the parameter $Re_F\delta^2$.}
    \label{fig:uf_rms_tot_vs_ReFDelta2}
\end{figure}

However, it is important to point out that, even though the r.m.s velocity profiles coincide for flows with similar values of $Re_F\delta^2$, the temporal characteristics of such flows differ. To corroborate this statement, the standard deviation of the temporal fluctuations of the horizontal velocities is calculated following Eq.~\eqref{eq:std_temp}. Fig.~\ref{fig:uf_rms_tot_vs_ReFDelta2} shows the total standard deviation of the horizontal flow fluctuations at the surface, $\sigma^f_U=\sqrt{[\sigma^f_u(\delta)]^2+[\sigma^f_v(\delta)]^2}$, as a function of $Re_F\delta^2$. The value of $\sigma_U^f$ is normalized with $\sigma_U$. 

As $Re_F\delta^2$ increases, the magnitude of fluctuations increases from values on the order of $ 10^{-4}$ to values close to unity, indicating that both standard deviations are becoming approximately equal (as $\langle u \rangle\approx\langle v \rangle\approx0$ and likewise $\langle u \rangle_t\approx\langle v \rangle_t\approx0$). Smaller fluctuations indicate a steady flow, and they increase in magnitude as the flow becomes more chaotic and time-dependent. In some instances, two flows with similar $Re_F\delta^2$ exhibit different magnitudes of the temporal fluctuations. Thus, despite having similar velocity profiles, the temporal behavior of each flow is different (one flow, for example, can be steady, while the other is time-dependent). Importantly, the appearance of temporal fluctuations coincides with the transition from the viscous to the inertial regimes. For the simulations with $\delta=0.1$, the transition occurs $Re_F\delta^2\approx 10$ as theoretically predicted for cases with $\delta\ll 1$. For the simulations with $\delta=0.3$ and 0.5, the appearance of the temporal fluctuations occurs at slightly higher values of $Re_F\delta^2$ ($Re_F\delta^2\approx 15$ and 20, respectively). 

\subsubsection{Vertical velocity}

\begin{figure*}
    \centering
    \includegraphics[scale=0.33]{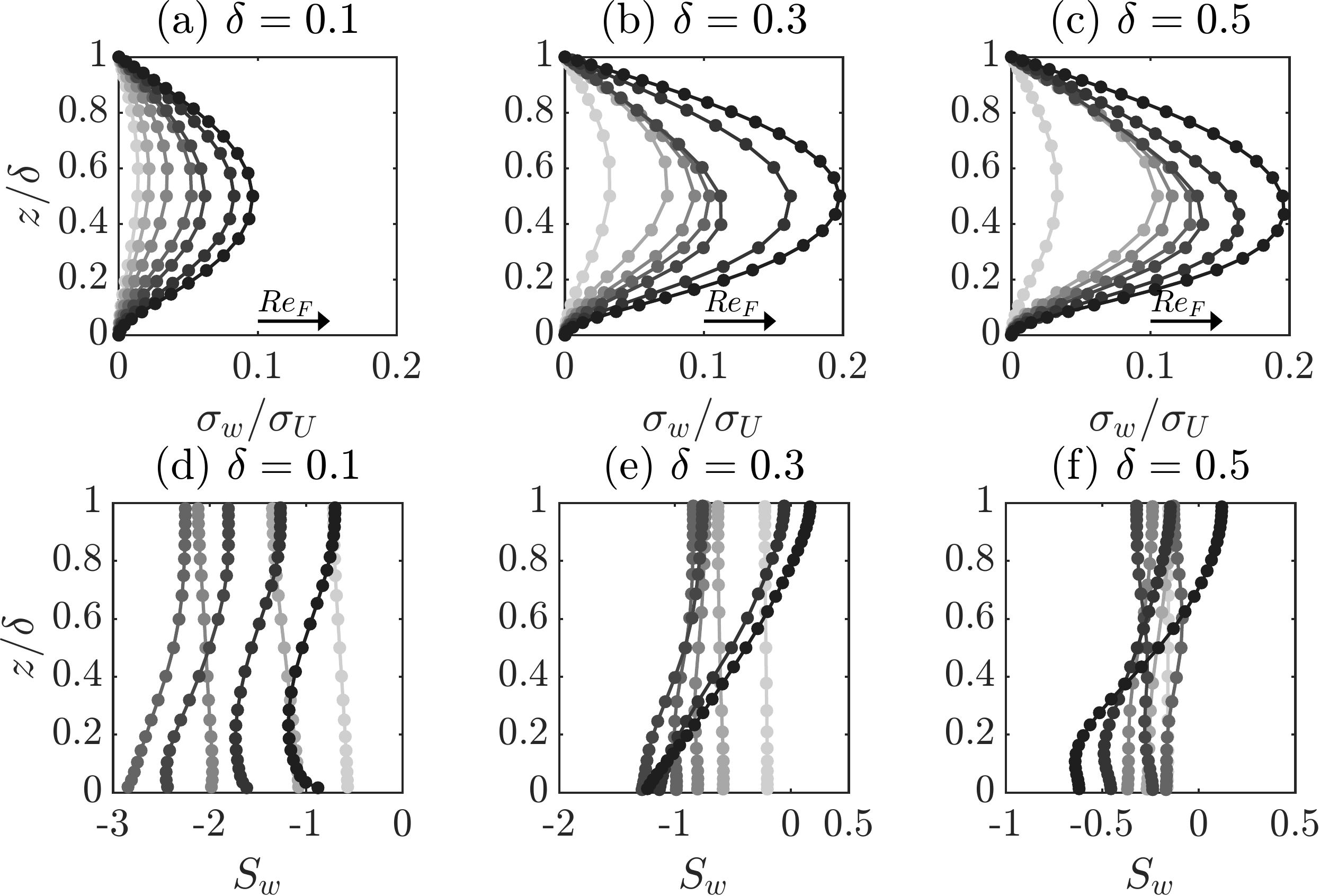}
    \caption{Vertical profiles of the normalized standard deviation of the vertical velocity $\sigma_w$ (panels a-c) and its skewness  $S_w$ (panels d-f) for all simulations. Each subfigure corresponds to a different $\delta$ value as indicated in their title. The arrow in the top panels indicates increasing $Re_F$ values. The grey color legend is the same as the one used in Figure \ref{fig:zprofiles_horizontal}.}
    \label{fig:zprofiles_vertical}
\end{figure*}

The mean vertical velocity $\langle w \rangle$ is zero for all depths and simulations because total upward flow must equal the total downward flow, as required by continuity. Fig.~\ref{fig:zprofiles_vertical}a-c shows the normalized vertical profiles of $\sigma_w=\langle w^2\rangle^{1/2}$ (equivalent to the r.m.s value since the mean is zero) for all simulations. These profiles have a maximum close to mid-depth, and they decrease to zero at the rigid top and bottom boundaries.

The vertical profiles of the skewness of the vertical velocity component $S_w$ are shown in Figs.~\ref{fig:zprofiles_vertical}d-f for all the $\delta$ values. This quantity gives an indication of the strength and of the occupied area of upward and downward motions. If the skewness is negative, there are strong downward motions covering a small area, and weaker upward motions covering a larger area. Conversely, positive skewness represents strong upward motions covering a small area and weaker downward motions covering a larger area  \cite{Moeng1990Vertical-VelocityLayer}. 

\begin{figure*}
    \centering
    \includegraphics[scale=0.5]{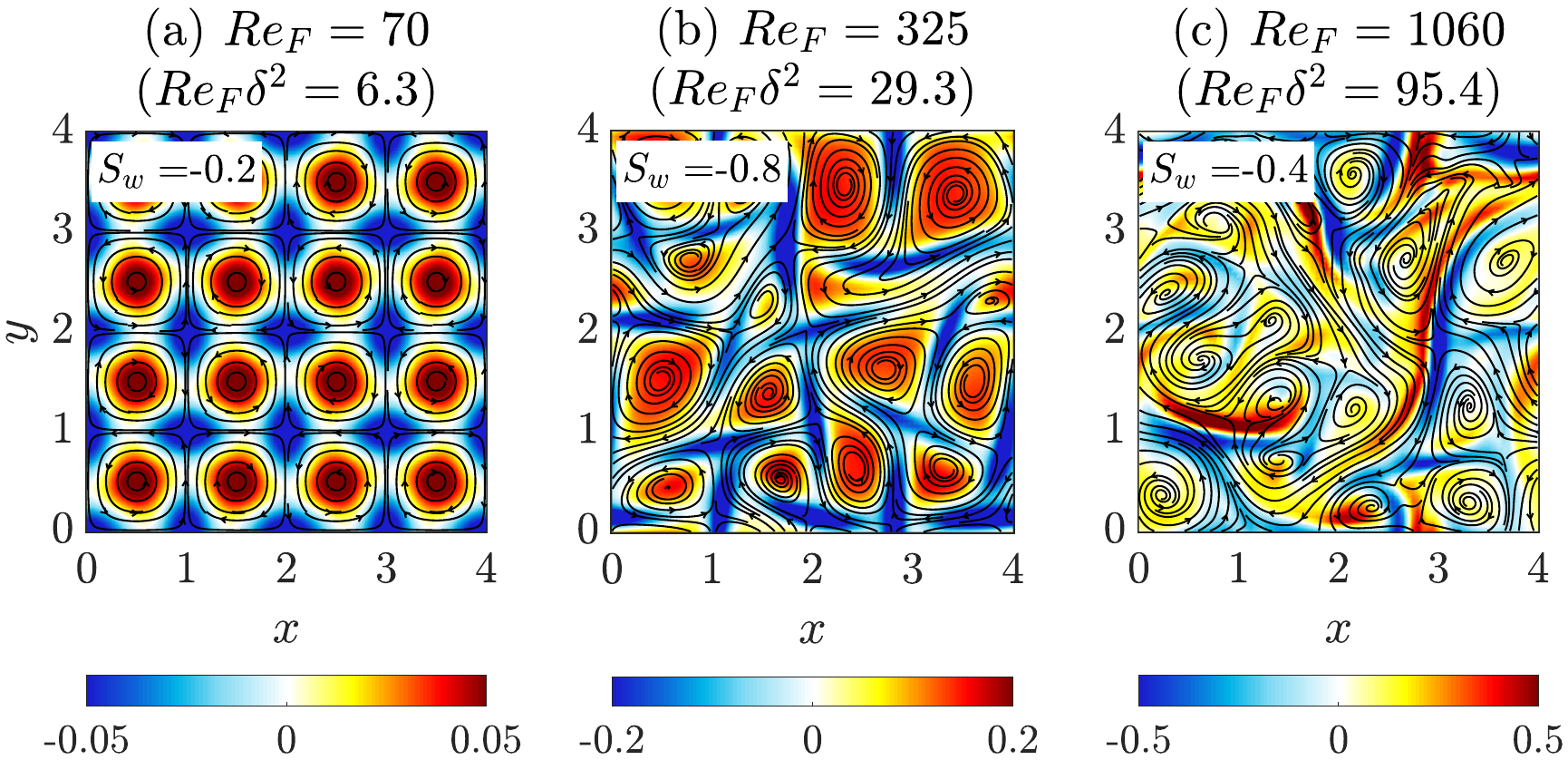}
    \caption{Snapshots of the normalized vertical velocity, $w/\sigma_U$, evaluated at mid-depth for shallow flows with $\delta=0.3$. The red (blue) color indicates upward (downward) flow. The black lines correspond to the flow lines tangential to the horizontal velocity. The inset rectangle gives the instantaneous value of vertical velocity skewness $S_w$.}
    \label{fig:wfields_deltas}
\end{figure*}

\begin{figure*}
    \centering
    \includegraphics[scale=0.5]{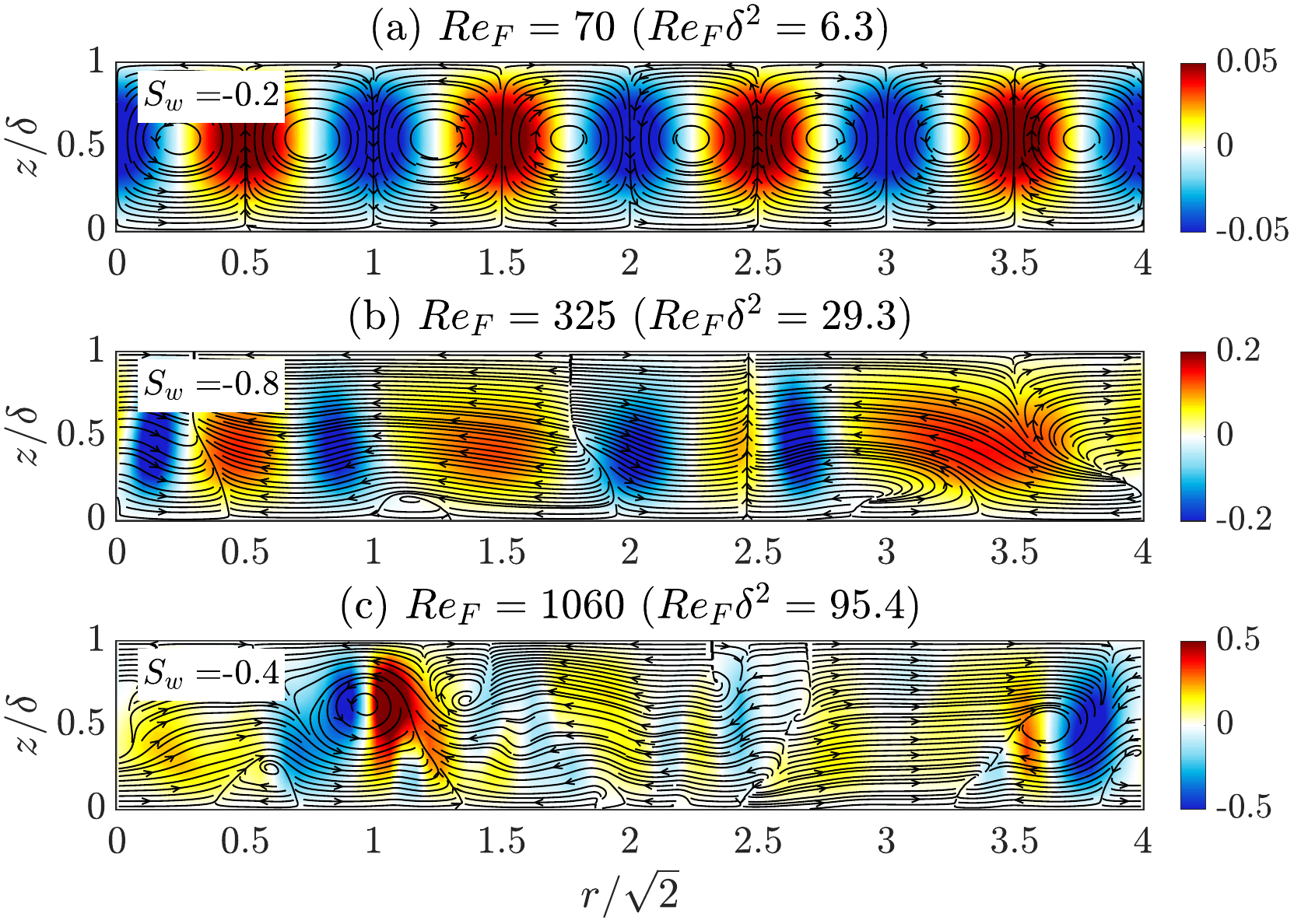}
    \caption{Vertical slices of the normalized vertical velocity, $w/\sigma_U$, for the shallow flows with $\delta=0.3$ along the $r,z-$plane, with $r$ denoting the line segment $x=y$. The red (blue) color indicates upward (downward) flow. The black lines correspond to the flow lines tangential to the velocity field in the $r,z-$plane. The inset rectangle provides the instantaneous value of vertical velocity skewness $S_w$ at mid-depth.}
    \label{fig:wvel_slices_deltas}
\end{figure*}

Unlike the skewness for the horizontal velocity components, which is equal to zero, the skewness for the vertical velocity is not. For most cases, $S_w< 0$, which indicates the above-described difference between upward and downward flows.  This is evident in snapshots of the vertical velocity field on the $x,y$-plane at mid-depth (Figs.~ \ref{fig:wfields_deltas}a and b). In addition, the uniformity of $S_w$ value across the fluid depth indicates that the updrafts and downdrafts extend over the whole layer, except near the rigid top boundary and the no-slip bottom wall. This can be seen in Figs.~\ref{fig:wvel_slices_deltas}a and b, where the vertical velocity field is presented in a $r,z-$slice (with $r$ the line segment $x=y$). 
 
The difference in these negative, uniform  profiles can be illustrated with the magnitude of the skewness. This magnitude decreases up to a minimum as the $Re_F$ value increases. For example, for the flows with $\delta=0.3$, the magnitude of the skewness goes from $S_w\approx -0.1$ for $Re_F=70$ ($Re_F\delta^2=6.3$) to a minimum around $S_w\approx -1$ for $Re_F=325$ ($Re_F\delta^2=29.3$). Hence, the skewness depends on the Reynolds number, indicating that the strength and area occupied by the updrafts and downdrafts are influenced by the flow conditions. For the smallest $Re_F$ (and $Re$) values, the updrafts and downdrafts tend to cover almost the same area and have a similar strength (Fig.~\ref{fig:wfields_deltas}a). In this case, well inside the viscous regime, the flow consists of a stationary square array of vortices, which is organized and maintained by the forcing \cite{Sommeria1986ExperimentalBox,Lauret2013, Forgia2022NumericalFlows}.  However, at $Re_F=325$ ($Re_F\delta^2=29.3$), the flow varies in time and becomes spatially disorganized as it transitions from the viscous to the inertial regime. In this case, the updraft areas are larger and the vertical velocities weaker in comparison with the strong and narrow downdrafts (Fig.~\ref{fig:wfields_deltas}b). Besides, the updrafts are occurring in extended, circular patches, while the downdrafts are concentrated in thin, elongated structures.

The minimum value of $S_w$ becomes more negative with smaller $\delta$ values. For example, at $\delta=0.1$, $S_w\approx -2.5$ for $Re_F\approx 1650$ ($Re_F\delta^2=16.5$) (Fig.~\ref{fig:zprofiles_vertical}d). This indicates that as the fluid layer becomes thinner, the asymmetry of vertical motions intensifies, with downdrafts strengthening and narrowing, while updrafts weaken and broaden. Conversely, for deeper layers, this asymmetry decreases. At $\delta=0.5$, $S_w$ remains around -0.25 for various $Re_F$ values (Fig.~\ref{fig:zprofiles_vertical}f).

\begin{figure*}
    \centering
    \includegraphics[scale=0.5]{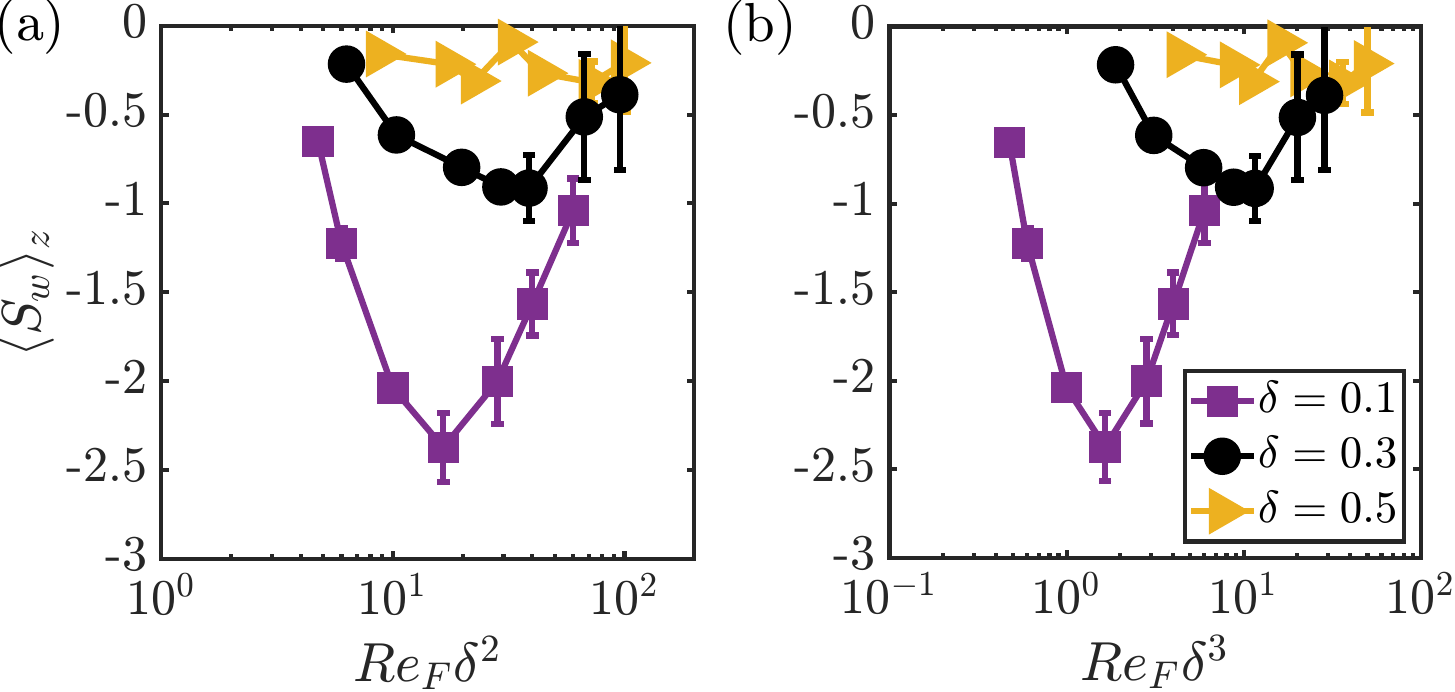}
    \caption{Mean value of the vertical velocity skewness, $\langle S_w\rangle_z$, as a function of (a) $Re_F\delta^2$ and (b) $Re_F\delta^3$. }
    \label{fig:Skw_vs_ReFDelta}
\end{figure*}

This variation of the skewness with $Re_F$ and $\delta$ is more apparent when it is plotted against the parameters $Re_F\delta^2$ and $Re_F\delta^3$ (Fig.~\ref{fig:Skw_vs_ReFDelta}). In particular, we plot the mean value of skewness, denoted as $\langle S_w\rangle_z$, where  $\langle\cdot\rangle_z=\frac{1}{\delta}\int_0^{\delta}\cdot\,\mathrm{d}z$ represents a vertical direction average. We observe that the layer thickness determines the minimum value of $S_w$, as previously mentioned. Moreover, this value is reached after the transition of the flow from the viscous to the inertial regime. For instance, this transition occurs at $Re_F\delta^2\approx10$ for $\delta=0.1$ (hence, $Re_F\delta^3\approx 1$), while the skewness minimum is found at $Re_F\delta^2= 16.5$ and $Re_F\delta^3= 1.65$. The same holds for $\delta=0.3$. Subsequently, the skewness increases towards zero as both parameters increase. 

\begin{figure*}[t]
    \centering
    \includegraphics[scale=0.33]{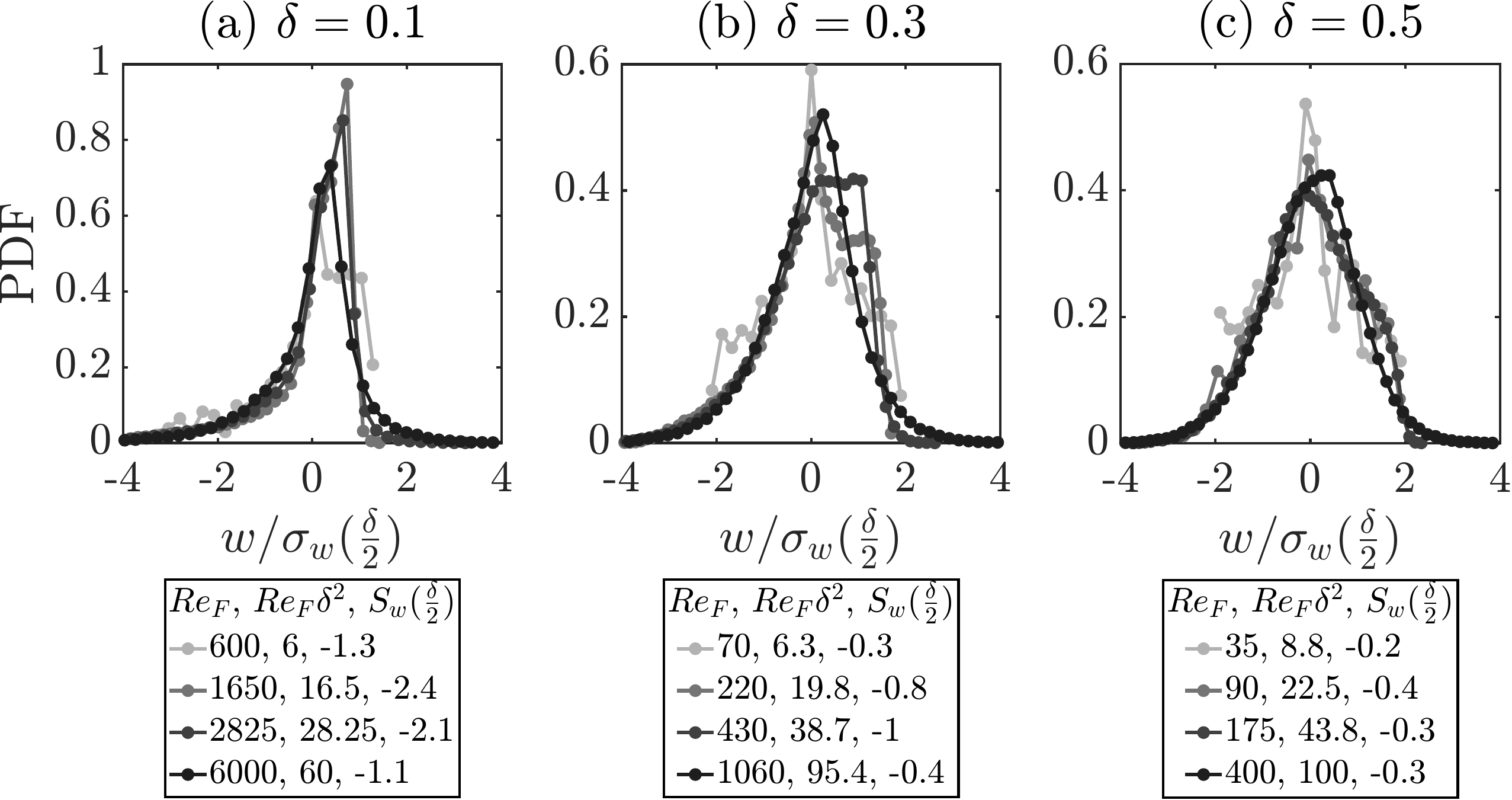}
    \caption{PDF of vertical velocities normalized by the standard deviation at mid-depth for selected simulations.  Each subfigure corresponds to a different $\delta$ value as indicated in their title.}
    \label{fig:PDFw_ALL}
\end{figure*}

Inspection of the PDFs of vertical velocities provides further insight into the transition in skewness. Fig.~\ref{fig:PDFw_ALL} shows the distributions for several $Re_F$ values at fixed $\delta$, constructed with the vertical velocities collected at mid-depth. For clarity, only representative cases of the viscous, transitional and inertial regimes are shown (see the $Re_F\delta^2$ values in the figure). In the viscous regime, the vertical flow exhibits more or less a similar symmetric distribution for the different values of $\delta$. As the flow transitions to the inertial regime, the distribution becomes strongly asymmetric (skewed), specially for the scenarios with $\delta=0.1$ and 0.3. In particular, the distribution exhibits a long tail for negative velocities on the left side and a limited range of positive velocities with high probabilities (occurrence) on the right side (Figs.~\ref{fig:PDFw_ALL}a and b). The occurrence of such positive velocities is more pronounced for the thinner layer ($\delta=0.1$), leading to larger values of $S_w$ in comparison with those found in $\delta=0.3$. This difference between the sides of the PDFs reflects the spatial characteristics of the upward and downward motions. Vertical velocities within the downdrafts are widely distributed, spanning a large range; while inside the updrafts, the vertical velocities are distributed more narrowly. As $Re_F\delta^2$ increases and the flow reaches deeper into the inertial regime, the range of positive velocities in the distribution extends, creating a long tail on its right side. At this point, the distribution becomes close to symmetric with two long tails on both sides, indicating similar spatial structures for both positive and negative velocities. This is evident in Fig.~\ref{fig:wfields_deltas}c, where both positive and negative velocities organize into thin, elongated filaments.

Finally, for large $Re_F\delta^2$ values (mostly in the inertia dominated regime), the profiles of skewness are no longer uniform in the vertical. For $\delta=0.3$ and 0.5, the skewness can even be weakly positive near the upper surface (for $z/\delta\gtrsim 0.8$), while it is negative for other depths. Recall that these flows consists of a boundary layer near the bottom and an almost inviscid interior (see Fig.~\ref{fig:zprofiles_horizontal}). While the larger negative values of $S_w$ reside close to the bottom, the positive values lie in the inviscid region indicating that intense and narrow updrafts exist in this region. In Fig.~\ref{fig:wvel_slices_deltas}c, swirling motions (i.e. horizontal vortical structures) are seen in the inviscid region above the bottom boundary layer, for example, in the regions around $r/\sqrt{2}=1$ and $r/\sqrt{2}=3.5$. They have intense positive and negative vertical velocities, i.e. strong updrafts and downdrafts. These vortices are reminiscent of the spanwise vortices that form at the front of propagating shallow dipolar vortices when the boundary layer is thinner than the total fluid depth (see \cite{Sous2004TurbulentLayer,Akkermans2008,Duran-Matute2010DynamicsVortices,Albagnac2014ADipole}).

All in all, the skewness of vertical velocities shows that there is an asymmetry in the distribution of updrafts and downdrafts in shallow flows. In particular, the weak updrafts cover a larger horizontal area compared to the intense, narrower downdrafts. Furthermore, the asymmetry depends on the vertical confinement, given by the aspect ratio, and the relative flow strength, given by the Reynolds number. 

\subsection{Connection between horizontal and vertical motions}

From the snapshots in Fig.~\ref{fig:wfields_deltas}, a noticeable relation is observed between the vertical velocities and the horizontal flow lines (which are lines locally tangential to the horizontal velocity field). For instance, in Fig.~\ref{fig:wfields_deltas}a, the upwellings are observed in regions characterized by closed flow lines, that is, where the vortices lie. These regions are referred to as elliptic regions. On the other hand, the downwellings occur in the regions where the flow lines converge in one direction and diverge in another, specifically in the spaces between the vortices. The regions are known as hyperbolic regions. This observed relation appears to persist even when the flow is distorted by inertia effects when the Reynolds number increases, as shown in Fig.~\ref{fig:wfields_deltas}b. However, this connection becomes less apparent for even larger $Re_F$ values (Fig.~\ref{fig:wfields_deltas}c). Previous studies on shallow flows have already explained and quantified the connection between the vertical flow and the elliptic and hyperbolic regions of the horizontal flow for isolated vortices  \cite{Kamp2012} and freely-decaying multiple-vortex flows \cite{Cieslik2010}. However, in contrast with such studies, our investigation focuses on forced flows and extends over a larger range of Reynolds number values. Therefore, it is essential to quantify the extent to which this relation is maintained within the considered range.

A diagnostic tool to distinguish the elliptic and hyperbolic regions within the horizontal flow is the so-called Okubo-Weiss (OW) parameter $Q$ \cite{Okubo1970HorizontalConvergences,Weiss1991TheHydrodynamics}. Given a 2D flow field $(u,v)$, this parameter is defined as

\begin{equation}
    Q=s_1^2+s_2^2-\omega^2, \label{eq:EulOW}
\end{equation}

\noindent where $s_1=\partial_x u-\partial_y v$, $s_2=\partial_x v+\partial_y u$ are the normal and shear strain, respectively, and $\omega=\partial_x v-\partial_y u$ is the vorticity. The elliptic regions are located where  vorticity dominates $(Q<0)$, whereas the hyperbolic regions are located where strain dominates $(Q>0)$. 

To investigate the correlation of the Okubo-Weiss parameter, which is a function of the horizontal flow, and the vertical velocity component, we follow the approach proposed by \citet{Cieslik2010}. First, the contribution of the OW parameter is divided in two components: $Q_+$ and $Q_-$, where $Q_+=Q$ if $Q>0$, and $Q_+=0$ otherwise, and $Q_-= Q$ if $Q \leq 0$, and $Q_-=0$ otherwise. Then, the correlation coefficient between the vertical flow $w$ and the strain-dominated part of the horizontal flow field, $Q^+$, is computed as:

\begin{equation}
    C_+=\dfrac{\langle Q_+w \rangle }{\sqrt{\langle Q_+^2 \rangle}\sqrt{\langle w^2\rangle}},
\end{equation}
where the averaging $\langle \dots \rangle$ is performed according to Eq.~\eqref{eq:average} and evaluated at $z/\delta=0.5$. Similarly, the correlation coefficient between the vertical velocities and the rotation-dominated part, $Q_-$, is obtained as:

\begin{equation}
        C_-=\dfrac{\langle Q_-w \rangle }{\sqrt{\langle Q_-^2 \rangle}\sqrt{\langle w^2 \rangle}}.
\end{equation}

\begin{figure*}[t]
    \centering
    \includegraphics[scale=0.5]{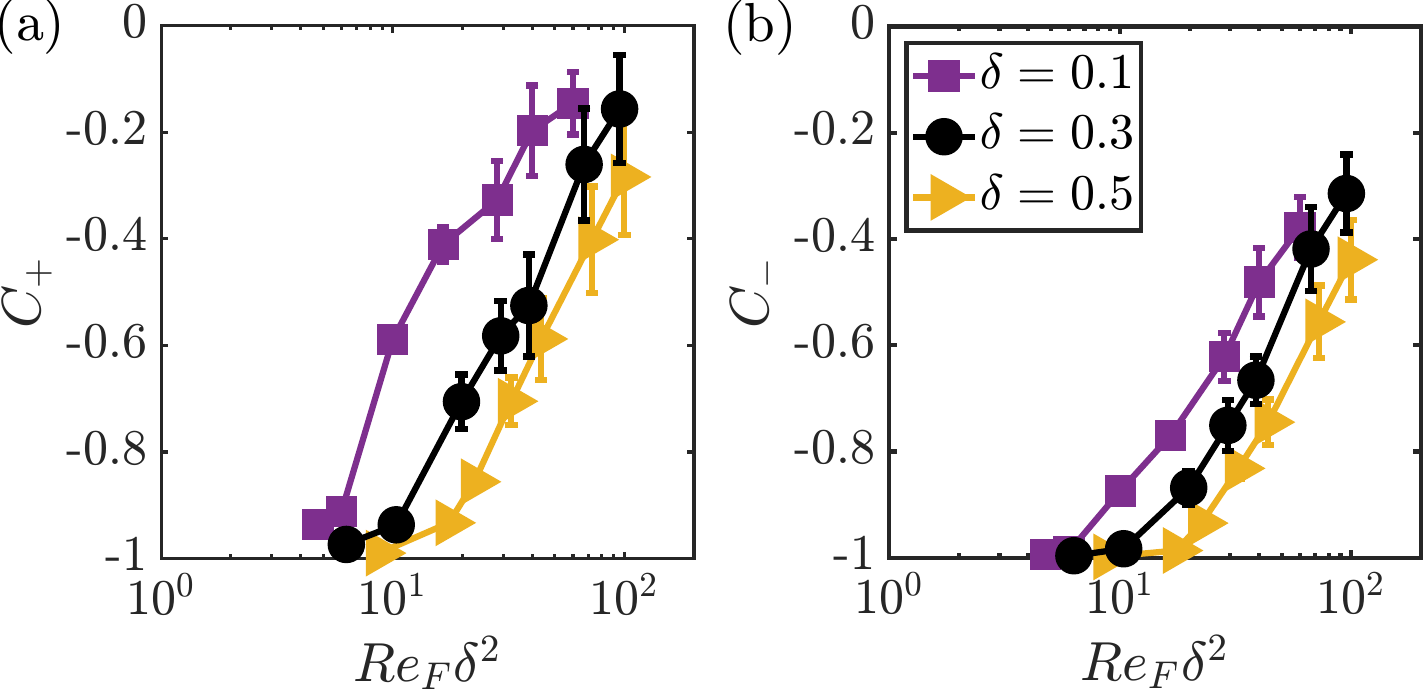}
    \caption{Correlation coefficients $C_+$ and $C_-$ as a function of the parameter $Re_F\delta^2$ for the three different aspect ratios $\delta$.}
    \label{fig:C1C2_vs_Re}
\end{figure*}

The correlation coefficients $C_+$ and $C_-$ are plotted against $Re_F\delta^2$ for the three values $\delta$ in Fig.~\ref{fig:C1C2_vs_Re}. In all cases, the coefficients are negative, supporting the notion that negative vertical velocities (downdrafts) are associated with strain-dominated regions and positive vertical velocities (updrafts) with rotation-dominated areas.  

The large magnitude of the coefficient $C_+$ ($C_-$) reveal that updrafts (downdrafts) are fully correlated with the strain- (vorticity-) dominated regions of the horizontal flow provided that the flow is within the viscous regime ($Re_F \delta^2\lesssim 10$). Previous studies have already shown that when the shallow flow is dominated by bottom friction, as occurs when $Re_F \delta^2$ is small, the upwellings and downwellings are determined by the spatial distribution (sign and magnitude) of the OW parameter \cite{Kamp2012}. However, this correlation weakens as $Re_F \delta^2$ increases. As for other previously discussed quantities, the transition occurs for larger values of $Re_F \delta^2$ for increasing $\delta$ values. Still, for some intermediate $Re_F \delta^2$ values ($50\gtrsim Re_F \delta^2\gtrsim 10$), there is a strong correlation between the direction of the vertical velocity and the strain- or vorticity-dominated regions. Typically, $0.6\lesssim |C_+|,|C_-|\lesssim 1.0$ for these cases.

\section{Vertical transport of particles}\label{sec:vert_transp}

To study the collective behaviour of particles in the vertical direction, first, the particle trajectories were obtained with the procedure described in section \ref{sec:LagPartTrack}. Here, we consider only the flows in the inertial regime ($Re_F\delta^2\geq 10$) with temporal fluctuations $\sigma_U^f/\sigma_U>10^{-1}$ (see Fig.~\ref{fig:uf_rms_tot_vs_ReFDelta2}). We neglect the simulations in the viscous flow regime because, as seen in the previous section, the flow is stationary, and hence, the behavior of the particles can be evaluated and interpret easily from the Eulerian velocity field. 

\subsection{Distribution of particles in the vertical direction}

\begin{figure*}[t]
    \centering
    \includegraphics[scale=0.5]{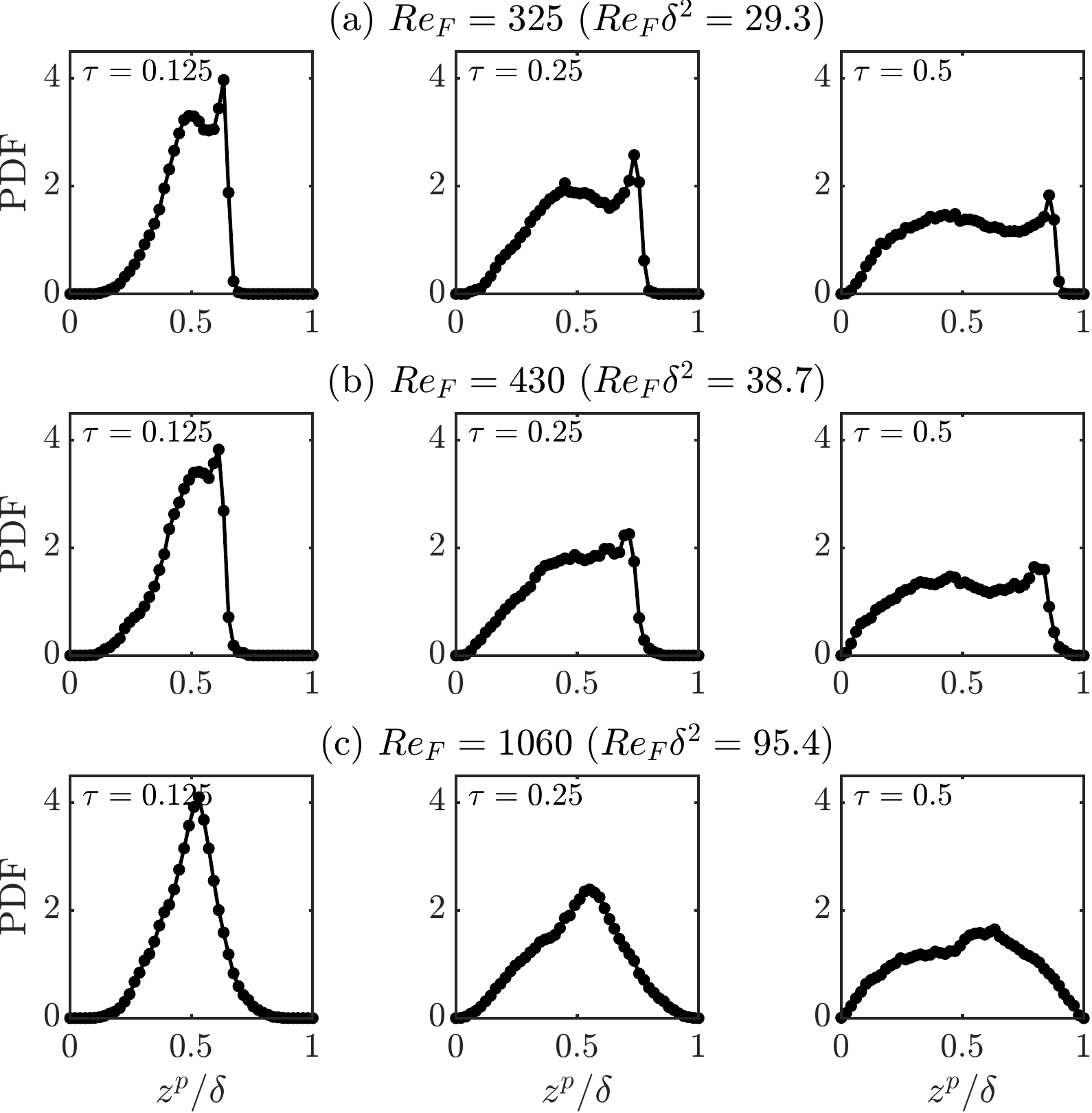}
    \caption{PDFs of the vertical particle position $z^p/\delta$ at different times $\tau$ for three simulations with $\delta=0.3$ and $Re_F=325$ (top row), 430 (middle row) and 1060 (bottom row).} 
    \label{fig:pdfzp_interm}
\end{figure*}

To elucidate the properties of vertical transport, we computed the probability density function (PDF) of the particle's vertical position $z^p/\delta$ at different times. The particles were initially deployed at mid depth ($z/\delta=0.5$). Fig.~\ref{fig:pdfzp_interm} shows the PDFs of $z^p/\delta$ at different times after release for the flows with $\delta=0.3$ and three different $Re_F$ values. The time $\tau$ is defined as the time $t$ normalized with $\sigma_w/\delta$ evaluated at mid-depth. This last quantity represents a typical time scale for a parcel to travel in the vertical over the fluid depth.

The distributions show that the particle cloud spreads over time from mid-depth towards both the bottom and the surface of the layer. Of particular interest is the difference in the shapes of the PDFs at their left and right sides. For example, consider the PDFs of the vertical distribution of particles for $Re_F=325$ and $Re_F\delta^2=29.3$ (Fig.~\ref{fig:pdfzp_interm}a). A significant peak is observed on the right side of the distribution. As time progresses, the peak reduces its height because the particles spread over the fluid layer, while it continues moving towards the surface as a front. Eventually, the particles are distributed homogeneously within the fluid layer and the peak disappears (not shown). It is evident that a portion of particles consistently upwell as a front during a significant amount of time (prior to the domain becoming vertically saturated). In contrast, no peak is observed on the left side of the distribution, representing the downwelling particles. Instead, a long tail gradually extends over time towards the bottom.

With a larger Reynolds number value ($Re_F=430$ and $Re_F\delta^2=38.7$), Fig.~\ref{fig:pdfzp_interm}b, the distributions no longer exhibit this peak on  the right side. However, a difference persists between the edges of the distributions. In particular, the right side appears as a front while the left side forms a long tail. If the Reynolds number is further increased ($Re_F=1060$ and $Re_F\delta^2=95.4$, Fig.~\ref{fig:pdfzp_interm}c), these differences vanish, and both sides of the distribution are characterized by long tails. Therefore, at this point, the particles upwell and downwell in a similar fashion, forming these long tails in the distribution that progressively extend towards both the surface and the bottom. 

The different behaviour of the left and right-hand sides of the distribution of the particle positions depict an asymmetry in the upward and downward vertical transport. This asymmetry is a direct consequence of the skewed distributions of vertical velocities. For instance, in the case with $Re_F=325$, the positive side of the distribution has a range of velocities with high probabilities that abruptly decrease around  $1.8\sigma_w$ (similar as shown in Fig.~\ref{fig:PDFw_ALL}b). Beyond this point, the probabilities for larger velocities become close to zero. The front is likely formed by particles with velocities around $1.8\sigma_w$  since only a few particles can exceed such velocity. In contrast, the vertical velocity distribution shows a long tail on the negative side, see Fig.~\ref{fig:PDFw_ALL}b. Due to this broad range of negative vertical velocities, the particles cover a broad range of distances within a given time. This results in the observed long tail on left side of the PDFs of the vertical position of the particles. However, there is still the possibility that particles change direction as they travel, and this will be further explored in Section~\ref{sec:clouds}.

Since the shape of the vertical velocity distributions changes with the Reynolds number,  the up-down asymmetry in transport changes in response. For $Re_F=1060$, for example, the distribution of vertical velocities is more symmetric ($S_w=-0.4$), see Fig.~\ref{fig:PDFw_ALL}b.  The distribution exhibits two large tails on the positive and negative sides. Under these circumstances, the spreading of particles must be symmetrical. In other words, a similar number of particles move upwards and downwards with vertical velocities of comparable magnitudes.

\begin{figure*}[t]
    \centering
    \includegraphics[scale=0.5]{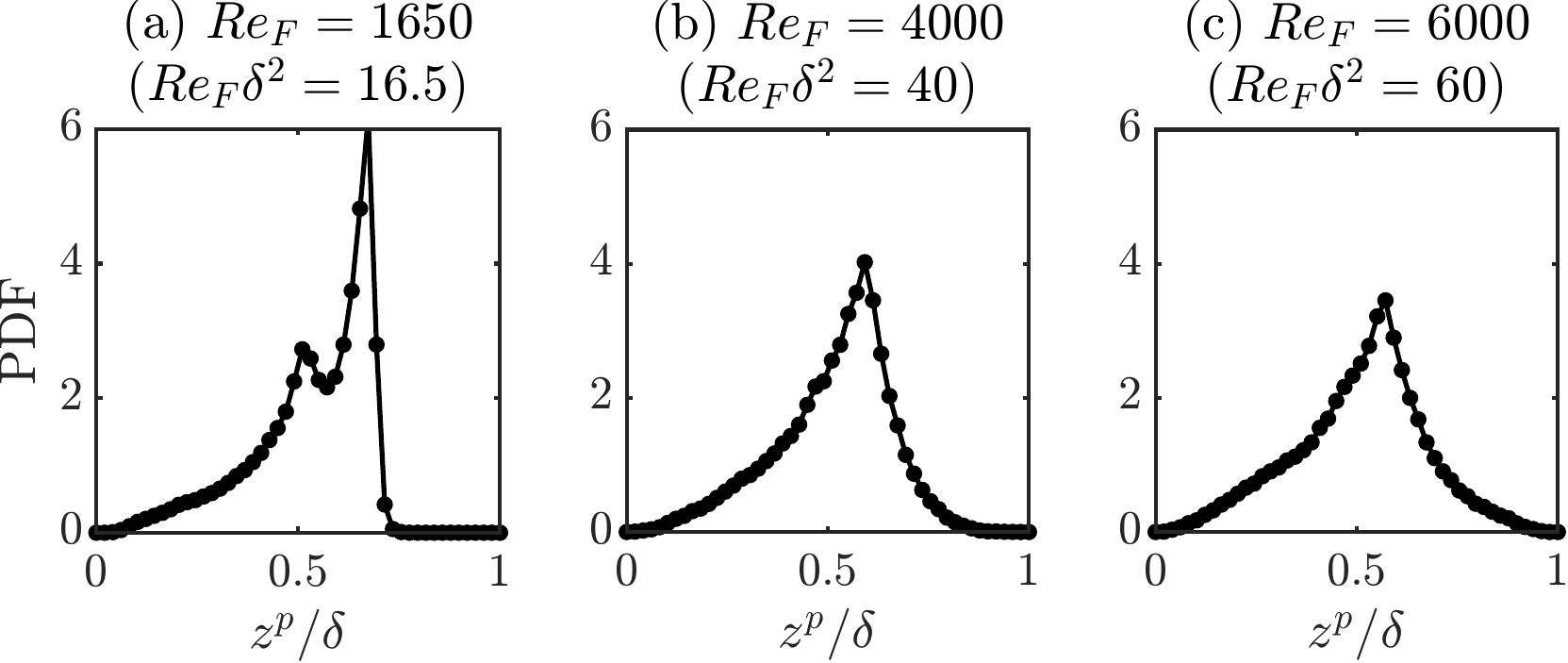}
    \caption{ PDFs of the vertical particle position $z^p/\delta$ at $\tau= 0.25$ for three simulations with $\delta=0.1$ and $Re_F=1650$, 4000 and 6000.}
    \label{fig:pdfzp_shallow}
\end{figure*}

Similar results are observed in the distributions of particle positions within the flows with $\delta=0.1$ (Fig.~\ref{fig:pdfzp_shallow}). For example, a front of ascending particles is also seen in the distributions for the scenario with $Re_F=1650$ (i.e. $Re_F\delta^2=16.5$) (Fig.~\ref{fig:pdfzp_shallow}a), similar to what is observed in the case with $Re_F=325$ and $\delta=0.3$ (i.e. $Re_F\delta^2=29.3$).  Similarly to the flows with $\delta=0.3$, the upward moving front changes with increasing values of $Re_F$ (Figs.~\ref{fig:pdfzp_shallow}b and c).  However, the peak behind the front is more pronounced for the shallower layer. This difference can be explained by considering the value of the vertical velocity skewness for the flow with $\delta=0.1$ ($S_w= -2.4$), which is larger than the skewness for the flow with $\delta=0.3$ ($S_w= -1$), and also supported by the vertical velocity PDFs shown in Fig.~\ref{fig:PDFw_ALL}a. This implies that the area covered by the positive velocities is larger in the flow with $\delta=0.1$. The larger coverage of positive velocities in the shallower flow results in more material being carried in the front.

\begin{figure*}[t]
    \centering
    \includegraphics[scale=0.5]{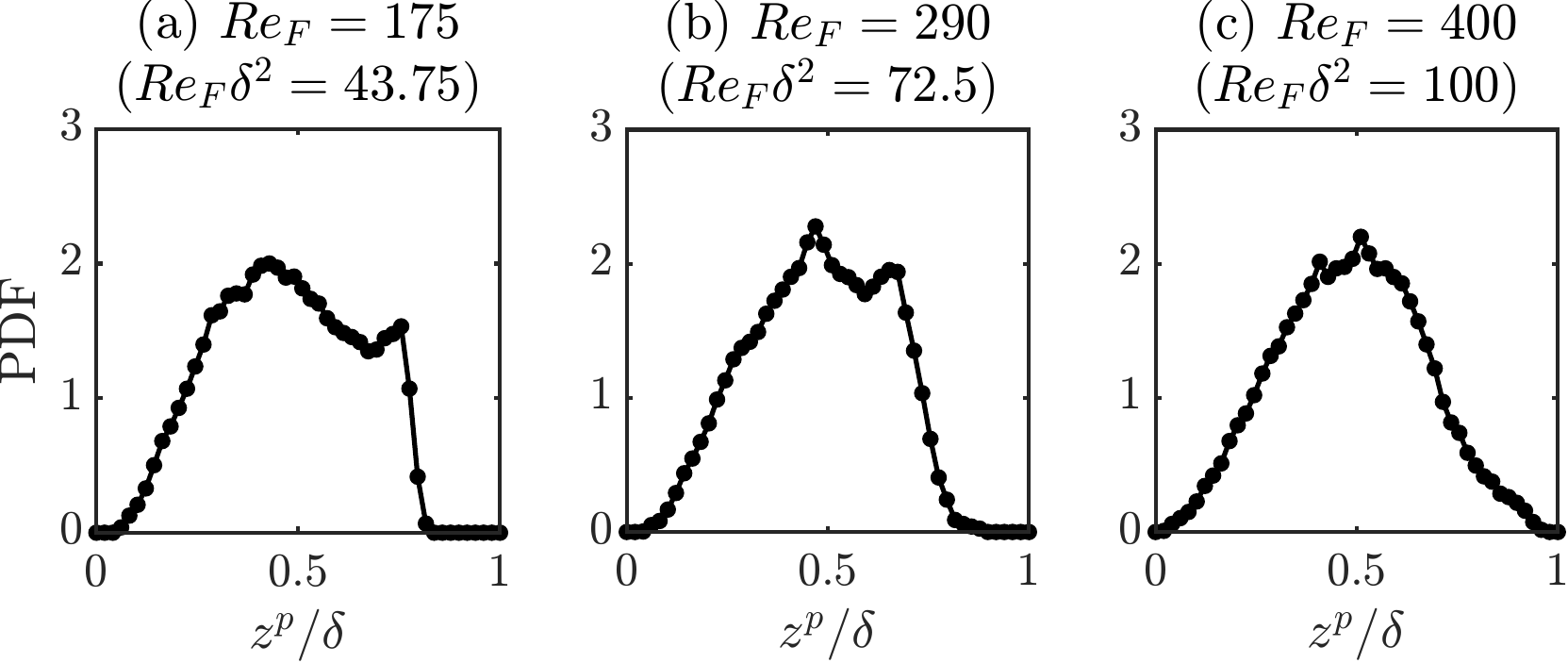}
        \caption{PDFs of the vertical particle position $z^p/\delta$ at $\tau= 0.25$ for three simulations with $\delta=0.5$ and $Re_F=175$, 290 and 400.}
   \label{fig:pdfzp_deep}
\end{figure*}

Hence, the asymmetry observed in the upward and downward transport is influenced by the fluid layer thickness, in addition to the Reynolds number. As the fluid layer becomes shallower (smaller $\delta$), this asymmetry increases. In contrast, it decreases if the layer becomes deeper (larger $\delta$). This is corroborated by considering the flows with a deeper layer ($\delta=0.5$). In such flows, the skewness remains negative, but relatively small ($S_w\approx -0.3$) for all the $Re_F$ values considered. In this context, the upwelling and downwelling regions are equally intense and occupy roughly the same area. As a consequence, the large peak of ascending particles is not observed in these flows (Fig.~\ref{fig:pdfzp_deep}), but clear signatures of the presence of the fronts are still present, see Figs.~\ref{fig:pdfzp_deep}a and b.

\subsection{Quantifying the asymmetry in vertical dispersion}

The aim of this section is to quantify the asymmetry in upward and downward spreading  observed in the PDFs of the vertical position of the particles. For this purpose, we focus on the edges of the distribution as they provide an indication of how the patch is expanding from mid-depth, where the particles are initially released, towards either the bottom or the surface. The selection of the edges is justified by the behaviour of the PDFs of the vertical particle position where, especially for shallower flows, a significant portion of particles is observed moving upwards in the left side of the distribution. 

In particular, we track in time the positions of the 5th and 95th percentiles of the distribution, denoted as $z_{5}$ and $z_{95}$, respectively.  The former indicates the downward spreading of the patch, while the latter provides the upward spreading. In addition, the 50th percentile position $z_{50}$ is obtained to track the center location of the cloud. Finally, we can obtain a velocity of propagation of each percentile position, denoted as $w_P$, by taking its temporal derivative, i.e., $w_P=\mathrm{d}z_P/\mathrm{d}t$, where $P=5$, $50$ and $95$. 

\begin{figure*}[t]
    \centering
    \includegraphics[scale=0.4]{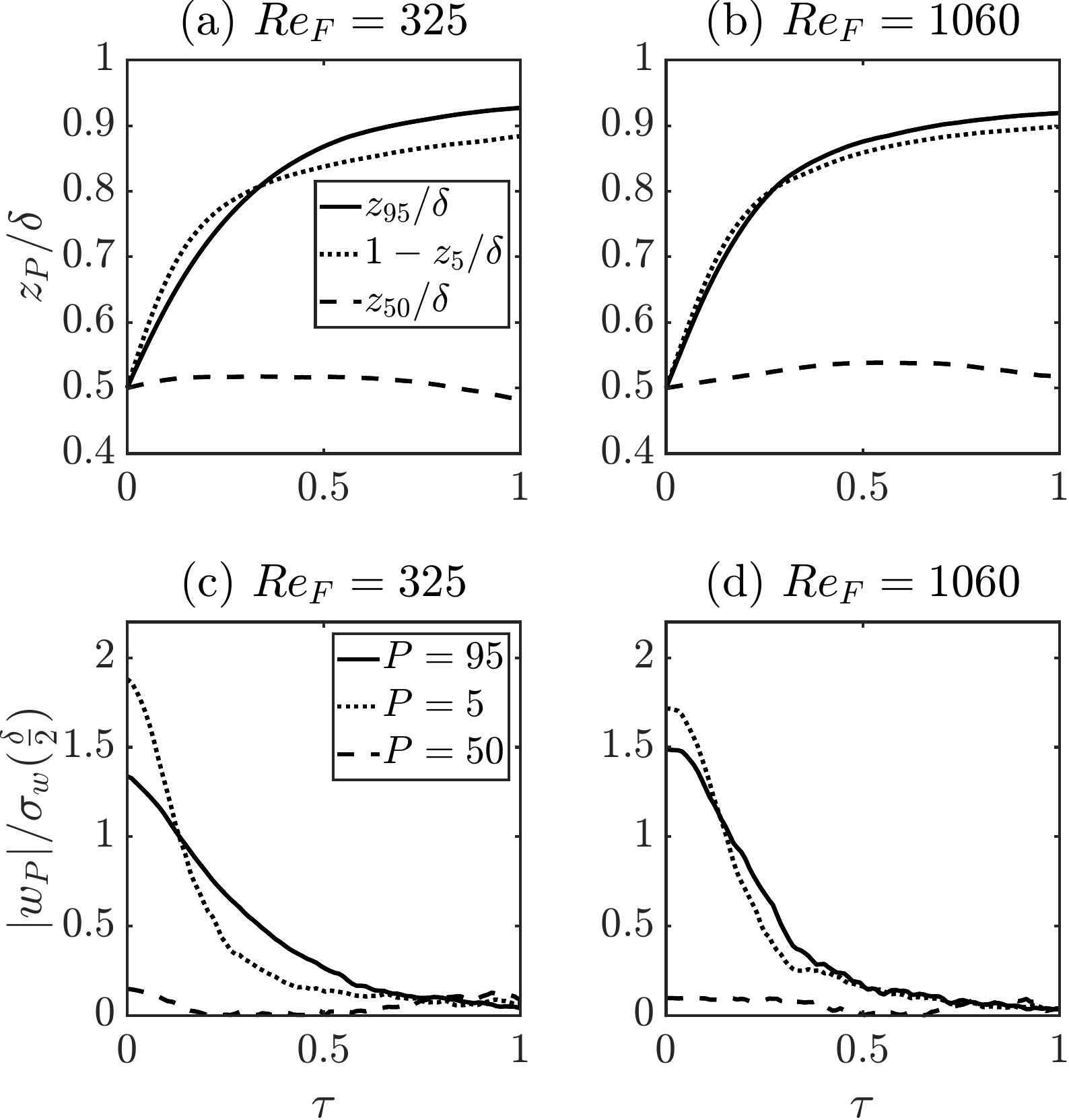}
    \caption{Temporal evolution of the position of the percentile $z_P$ (panels a and b) and the magnitude of its velocity $|w_P|$ (panels c and d) for two simulations with $\delta=0.3$ and $Re_F=325$ and 1060. The position $z_P$ is normalized with $\delta$, while the magnitude of $w_P$ is normalized with the standard deviation of the flow's vertical velocity $\sigma_w$ at $\delta/2$. For better comparison with the 95th percentile position, the 5th percentile position is plotted as $1 - z_5/\delta$. }
    \label{fig:zp_wp_d03}
\end{figure*}

Results of these calculations are presented in Fig.~\ref{fig:zp_wp_d03}, specifically for the case with $\delta=0.3$ and two $Re_F$ values. In Figs.~\ref{fig:zp_wp_d03}a and b, the temporal evolution of the normalized percentiles position $z_P/\delta$ is displayed for each $Re_F$ value. The position of 5th percentile is plotted as $1-z_5/\delta$ to facilitate the comparison with the 95th percentile position. Meanwhile, Figs.~\ref{fig:zp_wp_d03}c and d  show the curves of the magnitude of $w_P$ as a function of time for each $Re_F$ value, with the magnitude normalized by $\sigma_w$ evaluated at $\delta/2$.

For $Re_F=325$, different trajectories are observed for the percentiles $z_5$ and $z_{95}$ (Fig.~\ref{fig:zp_wp_d03}a), implying that particles at the edges of the cloud are advected upward and downward at different rates. This difference is also observed when examining the percentile velocities $w_5$ and $w_{95}$ (Fig.~\ref{fig:zp_wp_d03}c). The velocities exhibit different magnitudes from the initial time, with the velocity $w_5$ larger than $w_{95}$. Subsequently, both velocities decay to a value close to zero as the edges of the PDFs approach the surface and the bottom. However, the material is rapidly transported downwards to the bottom, faster than the material is advected upwards to the surface. Therefore, the velocity $w_5$ experiences a faster decay compared to $w_{95}$.  Note that the position of $z_{50}$ stays near the center of the layer and, as result, its velocity remains close to zero. Then, the difference in the upward and downward advection is unaffected by the shifting of the cloud towards either the surface or the bottom. 

These differences in $z_P$ and $w_P$ reduce with increasing Reynolds number value, as evidenced by the case of $Re_F=1060$ (Figs.~\ref{fig:zp_wp_d03}b and d). The behaviour of both percentile positions matches reasonably well during the entire evolution. This similarity is also reflected in the velocities of the percentiles. The asymmetry in the vertical transport disappears with large $Re_F$ values. In this scenario, the material is transported upward and downward at similar rates.

To quantify the degree of asymmetry in the vertical dispersion process, we introduce the asymmetry index $\mathcal{A}$ defined as:

\begin{equation}
    \mathcal{A}(\tau_0)=\dfrac{\overline{|z_{5}(\tau)-z_{50}(\tau)|}^\tau}{\overline{|z_{95}(\tau)-z_{50}(\tau)|}^\tau},
    \label{eq:asymm}
\end{equation}

\noindent where $\overline{\,\cdot\,}^\tau=\frac{1}{\tau_0}\int_0^{\tau_0}\cdot\,\mathrm{d}\tau$  is a temporal average over the trajectory, and $\tau_0$ is the time up to which the integration is performed. The quantity $\mathcal{A}(\tau_0)$ represents the ratio between the upward and downward mean distance from the center of the particle cloud up to time $\tau_0$. Distances are calculated relative to $z_{50}$ to account for any vertical drift of the particle cloud. If $\mathcal{A}=1$, the dispersion is symmetric, meaning that the upward and downward spreading are of the same order of magnitude. On the other hand, in the case of asymmetric dispersion, $\mathcal{A}$ becomes larger (or smaller) than one, which implies that the downward spreading is larger (or smaller) than the upward spreading. We define $\mathcal{A}$ in analogy to the definition of anisotropy usually employed in dispersion studies \cite{Ohlmann2019AnisotropyObservations}. In such context, the anisotropy is the ratio of spreading in two orthogonal directions. 

\begin{figure*}[t]
    \centering
    \includegraphics[scale=0.4]{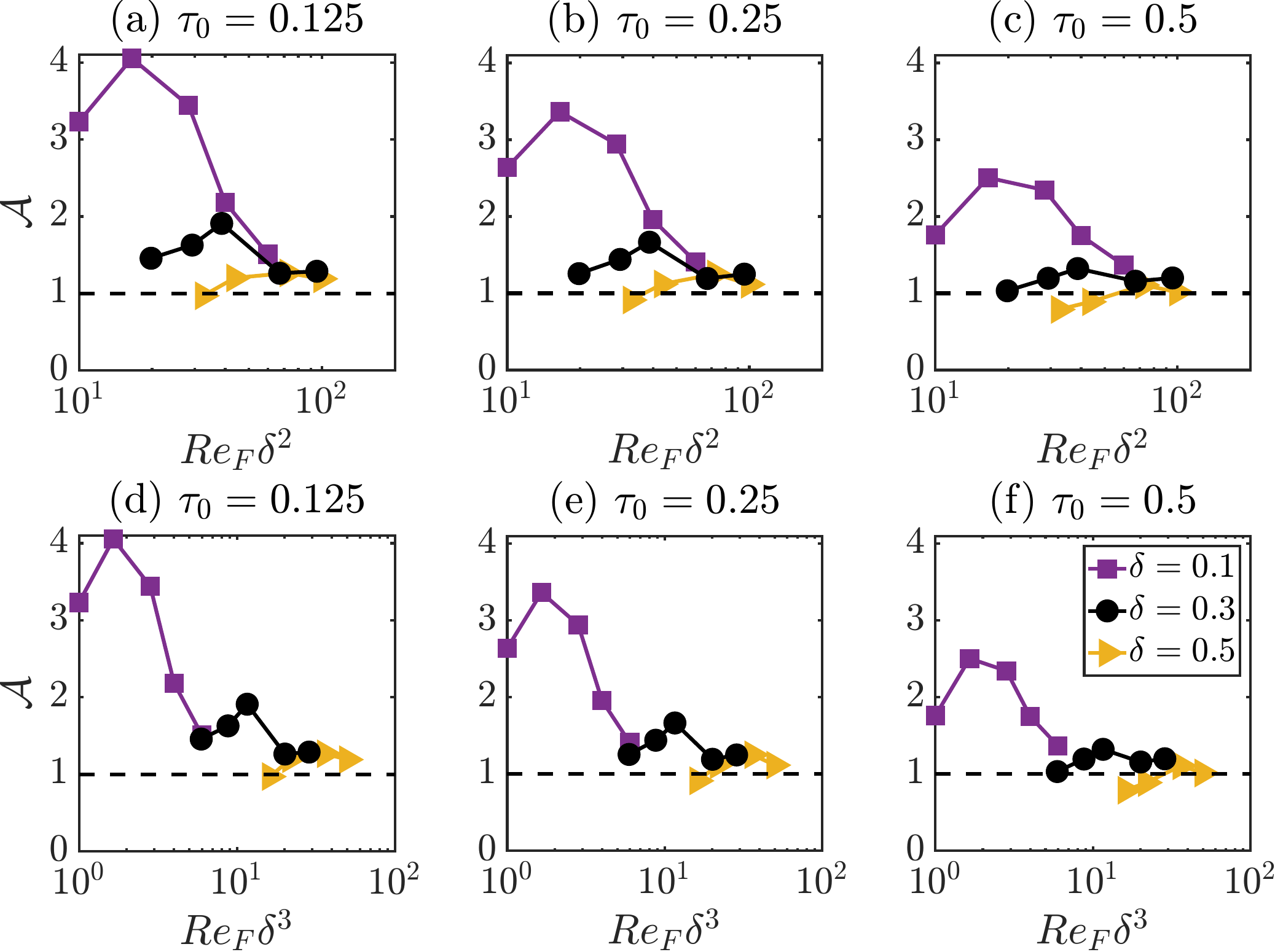}
    \caption{ Asymmetry index $\mathcal{A}$ as a function of the parameters $Re_F\delta^2$ (panels a-c) and $ Re_F\delta^3$ (panels d-f) for all shallow flows. The index has been evaluated in three different times $\tau_0$, as indicated in the panel titles. The horizontal dashed lines represent the index $\mathcal{A}=1$.}
    \label{fig:asymmetry_disp}
\end{figure*}

The index $\mathcal{A}$ is calculated for three times $\tau_0= 0.125$, 0.25 and 0.5. These times are selected to capture the entire evolution of the cloud, ranging from shortly after release to the moment when the layer is nearly covered with particles. Fig.~\ref{fig:asymmetry_disp} shows the indices $\mathcal{A}$ for these three times as a function of the parameters $ Re_F\delta^2$ (panels a-c) and $ Re_F\delta^3$ (panels d-f). In general, $\mathcal{A}\gtrsim1$, indicating the previously described asymmetry between downward and upward dispersion. 

The behaviour of $\mathcal{A}$ against $Re_F\delta^2$ further supports the previous observations regarding the PDFs of particle’s vertical position. On the other hand, the spreading becomes symmetric (i.e., $\mathcal{A}$ approaches unity) with increasing values of $Re_F\delta^2$ (Fig.~\ref{fig:asymmetry_disp}a-c). Furthermore, spreading exhibits more pronounced asymmetry in the shallower layer, while it tends to be more symmetric as the layer deepens. These two trends are more evident when the asymmetry is plotted as a function of $Re_F\delta^3$ (Fig.~\ref{fig:asymmetry_disp}d-f), as explained in the next section. This overall behaviour is maintained up to $\tau_0=0.5$ (Figs.~\ref{fig:asymmetry_disp}c and f) but with smaller values of $\mathcal{A}$ and slightly distorted. This is because the particles have saturated the domain, and both positions of the percentiles are affected by the confinement of the cloud of particles. It is worth noting that the behaviour of $\mathcal{A}$ closely resembles that of $-\langle S_w\rangle_z$ (Figs.~\ref{fig:Skw_vs_ReFDelta}a and b), reflecting the direct influence of the vertical flow (and its distribution) in the early spreading.

The index $\mathcal{A}$ confirms quantitatively the existence of an asymmetry in the vertical dispersion. Such asymmetry is a direct outcome of the skewed distributions of vertical velocities, and depends on the fluid layer thickness and the flow strength (as the skewness of those distributions). However, the Lagrangian metrics used so far are obtained from the entire cloud of particles, which is influenced by both horizontal and vertical velocities. The next step is to isolate the particle clouds moving exclusively upward or downward, and analyze the resulting Lagrangian statistics to understand additional underlying mechanisms.

\subsection{Motion of clouds with upwelling and downwelling particles }\label{sec:clouds}

To get further insight into the asymmetry in the vertical dispersion, we divide a posteriori the particles at every time $\tau$ into two groups: 1) those that move downward from the moment of release until time $\tau$, and 2) those that move upward from the time of release until time $\tau$. Consequently, particles that have changed their direction of motion at least once between the time of release and time $\tau$ are excluded from both groups. 

\begin{figure*}[t]
    \centering
    \includegraphics[scale=0.49]{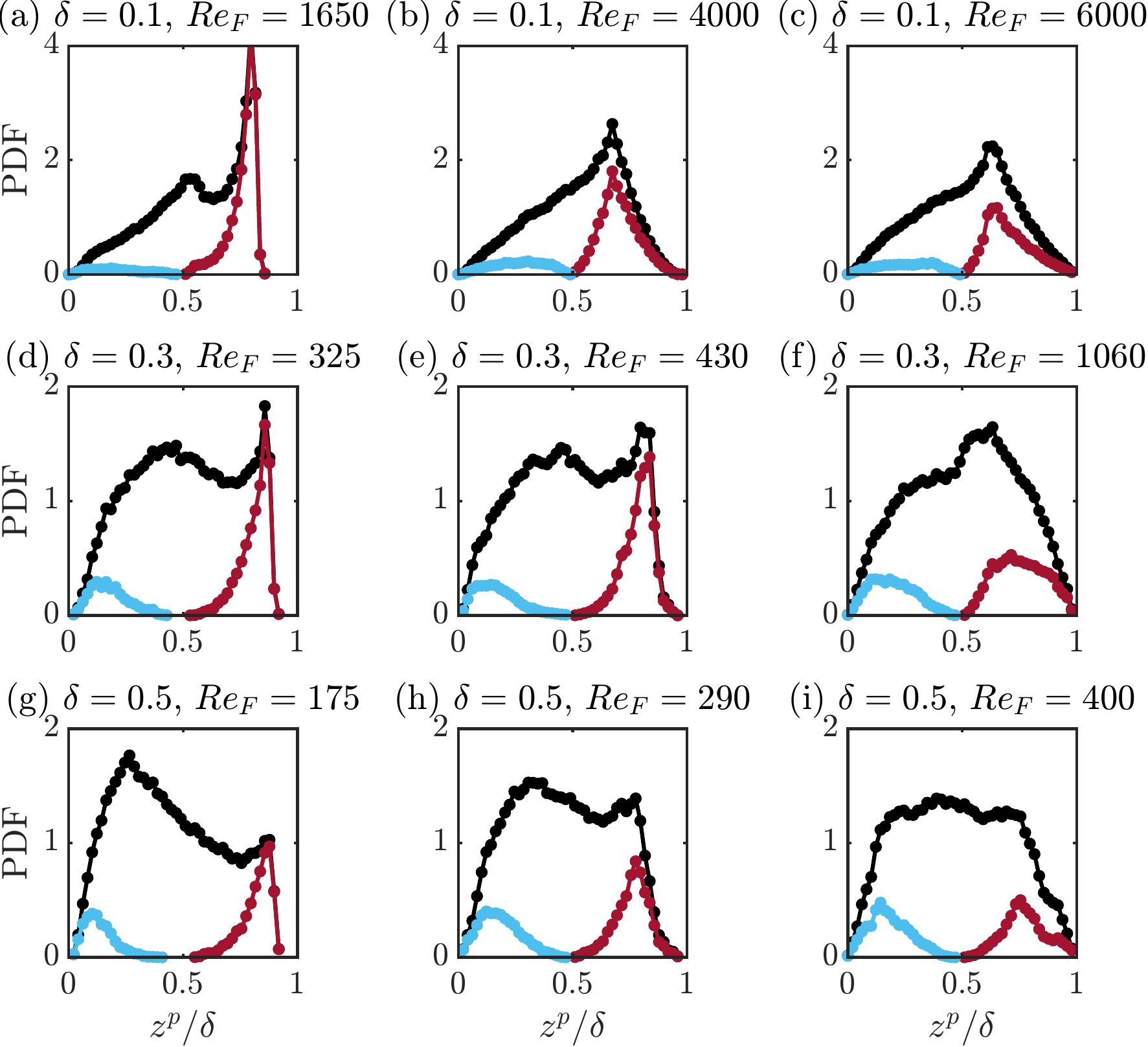}
    \caption{PDFs of the vertical particle position $z^p/\delta$ at $\tau= 0.5$ for different simulations. The panel titles specify the values of $Re_F$ and $\delta$ for each simulation. Each panel shows the PDFs constructed with the particles that maintain, until this time, a positive (red line and filled circles) and a negative (blue line and filled circles) vertical velocity. The black line (and filled circles) represents the PDF for all particles.}
    \label{fig:pdf_zp_all_tot_up_down}
\end{figure*}

We calculate the PDF of the vertical position $z^p/\delta$ for the particles belonging to each group. Fig.~\ref{fig:pdf_zp_all_tot_up_down} displays the distributions of these particles at $\tau= 0.5$, plotted alongside the PDFs for all particles, for various $Re_F$ values and the three different aspect ratios $\delta$. Even if the particles maintain their direction of propagation (i.e. they belong to one of the two groups) not all of them travel the same distance in a given time due to the difference in speed of the particles. Of particular interest is the difference in the shapes of the PDFs of particles belonging to one of the two groups because it shows the asymmetry in the upward and downward vertical spreading. In particular, the peak on the right-hand side of some of the total distributions (represented by the black dotted lines) is mainly formed by particles that have always moved upward (see Figs.~\ref{fig:pdf_zp_all_tot_up_down}a and d). In contrast, the left tail of the distribution is mainly formed by the particles that have always moved downward. Therefore, the  edges of the total distributions consist of particles that retain their original vertical velocity from the start. As the value of the Reynolds number increases, the shapes of the PDFs of both groups of particles become similar. For instance, in Fig.~\ref{fig:pdf_zp_all_tot_up_down}f, the distributions for the simulation with $Re_F=1060$ and $\delta=0.3$ show closely identical shapes for the upwelling and downwelling particles, without a prominent peak in the distribution of upwelling particles. Similar observation can be made in the simulation with $Re_F=400$ and $\delta=0.5$ (Fig.~\ref{fig:pdf_zp_all_tot_up_down}i).  

\begin{figure*}[t]
    \centering
    \includegraphics[scale=0.33]{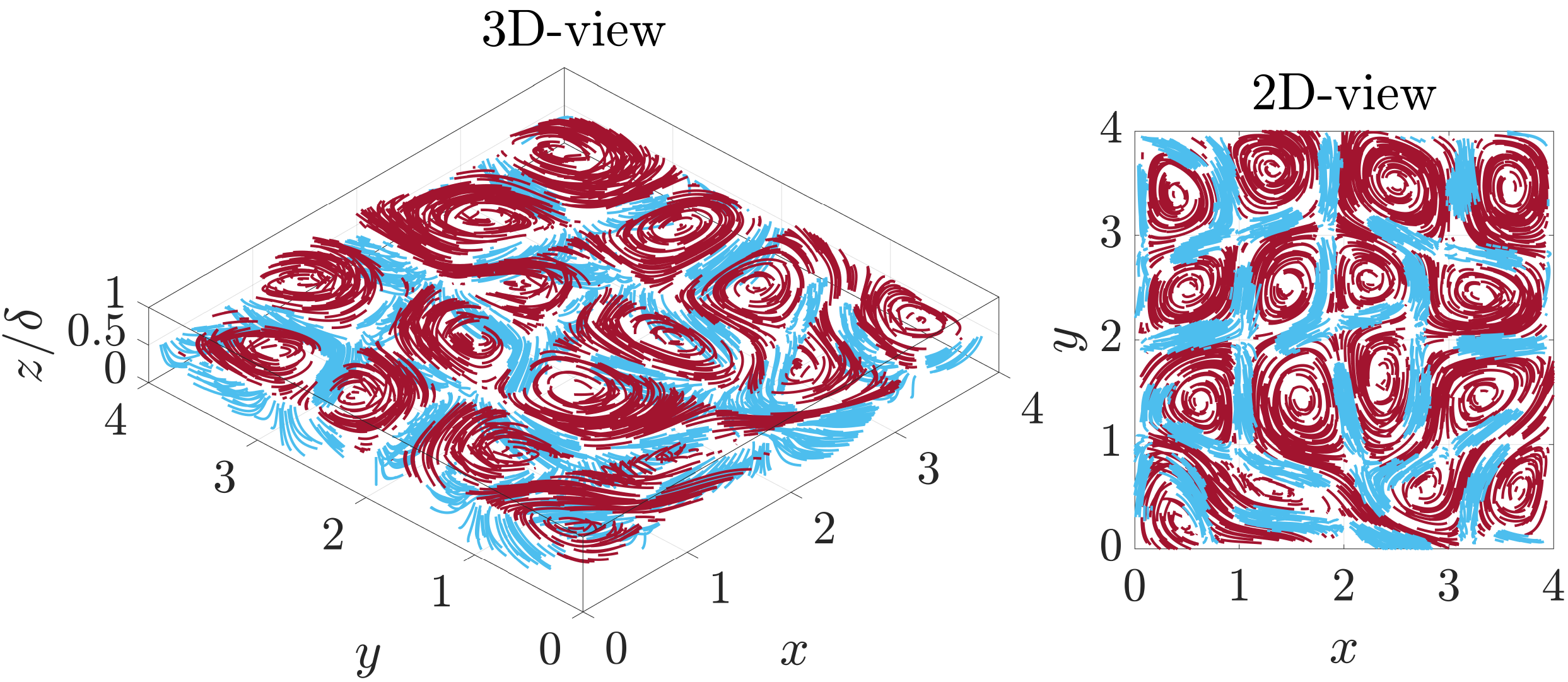}
    \caption{Trajectories of particles released  within the flow with $Re_F=1650$ and $\delta=0.1$. The particles are coloured according to their initial vertical velocity: red for particles with positive vertical velocity and blue for particles with negative vertical velocity.}
    \label{fig:particle_traj}
\end{figure*}

The different shapes of the distributions for the upwelling and downwelling particles arise not only from the distributions of vertical velocities but also from how these velocities are related to the horizontal flow structures. For example, Fig.~\ref{fig:particle_traj} shows the trajectories of particles released within the shallow flow with $Re_F=1650$ and $\delta=0.1$. The trajectories and particles are coloured using the partition described earlier. The particles coloured in red move upward mostly following horizontal circular paths while ascending (swirling motion) since they are found in vorticity-dominated regions (i.e. vortices). In contrast, the blue particles travel downwards in narrow streams in between the vortices (where the strain-dominated regions are found). 

\begin{figure*}[t]
    \centering
    \includegraphics[scale=0.33]{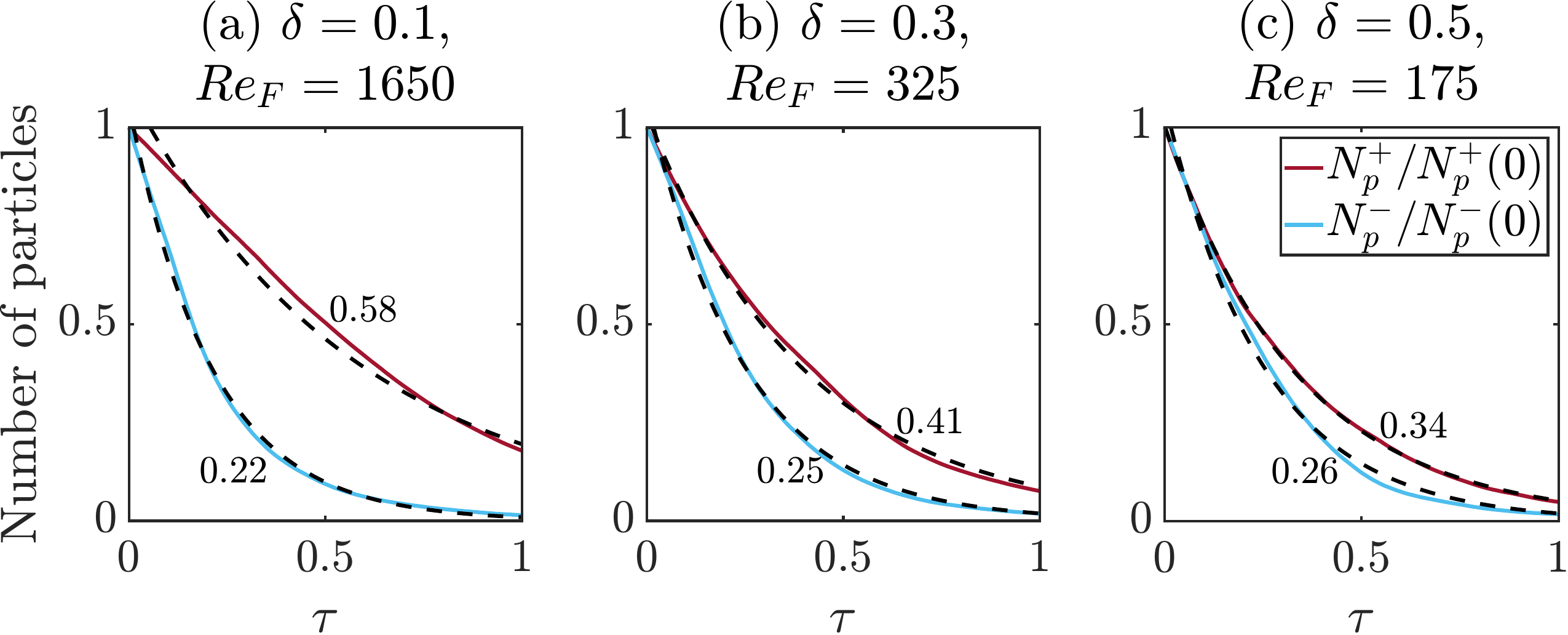}
    \caption{Temporal evolution of the number of particles retained by the updrafts ($N_p^+$) and retained by the downdrafts ($N_p^-$) for three different simulations (as the title of each panel indicates). The curves are normalized with their initial values. The dotted black lines represent exponential fits with a characteristic decay time $\tau_D$ corresponding to the value next to each fit.}
    \label{fig:numparticles_examples}
\end{figure*}

Moreover, the PDFs of the upwelling and downwelling particles (Fig.~\ref{fig:pdf_zp_all_tot_up_down}) reveal that less particles from the group of particles moving upwards change sign and are discarded.  We quantify the number of particles $N_p$ that remain inside the upwelling cloud, denoted as $N_p^+$, and the downwelling cloud, denoted as $N_p^-$, as a function of time.  Fig.~\ref{fig:numparticles_examples} displays the time evolution of $N_p^+$ and $N_p^-$, normalized with their initial value, for three different simulations. Besides, the curves are fitted with exponential functions of the form $\exp(-\tau/\tau_D)$ up to time $\tau=1$, where $\tau_D$ is a characteristic decay time.

The curves of $N_p^+$ and $N_p^-$ monotonically decay over time as the particles change their initial vertical velocity by moving to an area with vertical velocity with opposite sign. The curve for $N_p^-$ decays faster than that for $N_p^+$. This difference is more pronounced for flows close to the transition between the viscous and inertial flow regimes, where the particles remain in updrafts for longer times in comparison with those in downdrafts. Notice that particles in a stationary flow, such as in the viscous flow regime, would not change their vertical direction, unless they would reach the bottom or surface. Outside the viscous regime, some particles also alter their vertical direction only upon reaching a boundary, but this  occurs for $\tau \gtrsim 0.5$.

\begin{figure*}[t]
    \centering
    \includegraphics[scale=0.4]{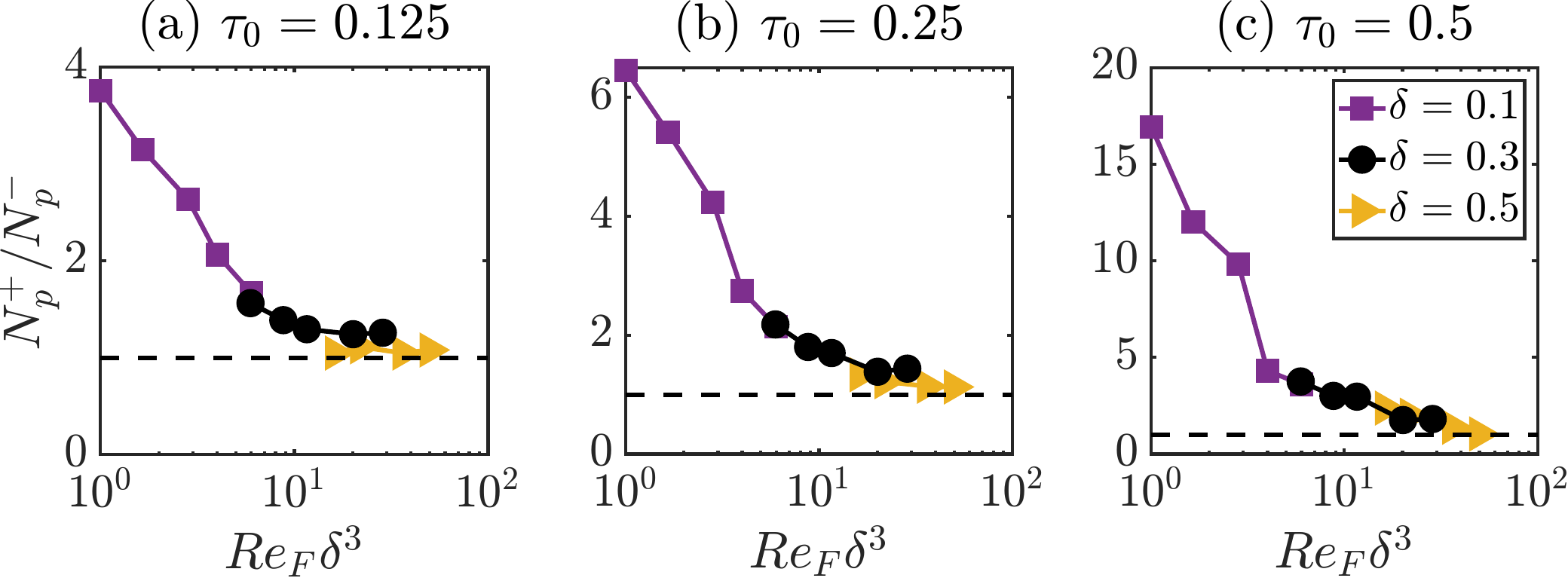}
    \caption{Ratio $N_p^+/N_p^-$ as a function of the parameter $Re_F\delta^3$ for different simulations. The ratio has been computed at different times $\tau_0$ as indicated by the title of each panel. The horizontal dashed line represents the ratio $N_p^+/N_p^-=1$.}
    \label{fig:numparticles_ratio}
\end{figure*}

The ratio $N_p^+/N_p^-$ is computed for all the simulations. Fig.~\ref{fig:numparticles_ratio} shows this ratio, obtained at three different times $\tau_0= 0.125,$ 0.25 and 0.5, plotted against the parameter $Re_F\delta^3$. This parameter was chosen because, as mentioned in Section \ref{subsec:flow_regimes}, the strength of the secondary vertical motions in shallow monopolar and dipolar vortices have been shown to scale with this parameter \cite{Duran-Matute2010DynamicsVortices,Duran-Matute2010ScalingFlows}.  A collapse of the curves is observed for $\tau_0=0.5$ and $\tau_0=0.25$, and less clearly for $\tau_0=0.125$ (due to a slight offset of the values with $\delta=0.5$). For $Re_F\delta^3\lesssim 10$, the value of the ratio is clearly larger than unity, indicating a greater retention of particles by updrafts compared to downdrafts.  However, for $Re_F\delta^3> 10$, the value of $N_p^+/N_p^-$ tends to unity, as the updrafts and downdrafts retain similar amount of particles and the asymmetry between upward and downward transport is reduced.

\begin{figure}[t]
    \centering
    \includegraphics[scale=0.45]{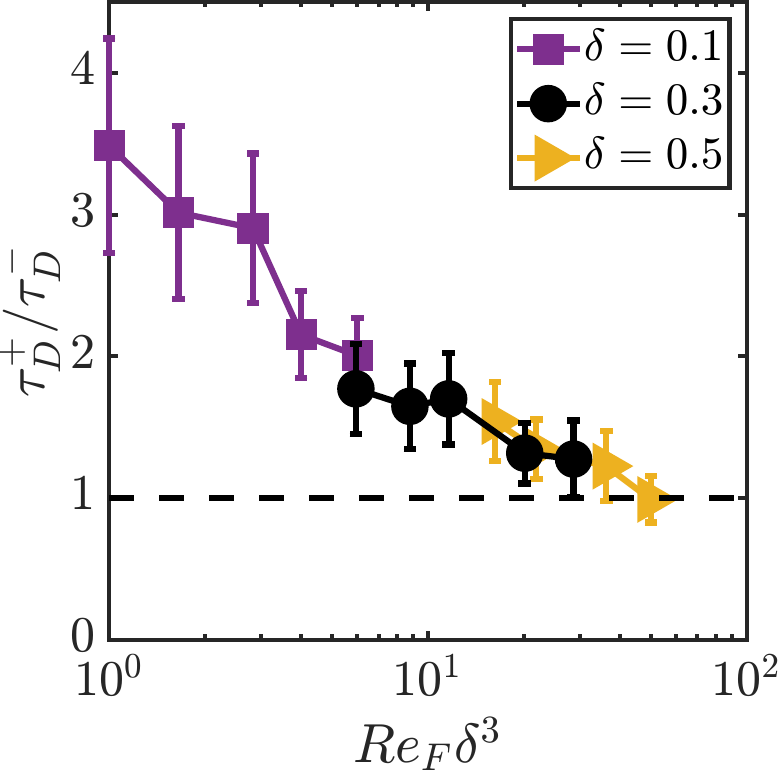}
    \caption{Ratio $\tau_D^+/\tau_D^-$ as a function of the parameter $Re_F\delta^3$ for different simulations. The horizontal dashed line represents the ratio $\tau_D^+/\tau_D^-=1$.}
    \label{fig:decaytimes_ratio}
\end{figure}

The values of $\tau_D^+$ and $\tau_D^-$, the characteristic decay times of $N_p^+$ and $N_p^-$ (see Fig.~\ref{fig:numparticles_examples}), are obtained for all the simulations, and combined into the ratio $\tau_D^+/\tau_D^-$, which is plotted against $Re_F\delta^3$ (Fig.~\ref{fig:decaytimes_ratio}). As result, the three curves, each one corresponding to a different $\delta$ value, almost connect forming a single curve. The trend is similar to the one observed in Fig.~\ref{fig:numparticles_ratio}. That is, for $Re_F\delta^3 \lesssim 10$, the ratio is larger than one because the updrafts retained more particles than the downdrafts. Meanwhile, for $Re_F\delta^3> 10$, the ratio is closer to unity, indicating comparable particle retention by both updrafts and downdrafts.

\section{Discussion and conclusions}\label{sec:discu_conclu}

The asymmetry in the vertical dispersion of particles can be understood in terms of the Eulerian characteristics of the simulated flows. The first feature is the asymmetric (skewed) distribution of their vertical velocities: strong downdrafts take place in smaller regions whereas weak updrafts occur in larger areas.  Similar distributions of the vertical flows have been observed in other shallow flows \cite{ Cieslik2010, Akkermans2012ArraysDispersion, Branson2019Three-dimensionalityWakes, Gargett2004LangmuirSeas}. Consequently, when particles are carried by such flows, their vertical spreading exhibit asymmetry as updrafts transport larger amount of particles but slowly, while downdrafts rapidly advect the remaining particles.

The second feature is the relation between the vertical velocities and the horizontal flow’s structures. This relation was also previously observed and quantified in experiments and simulations in a shallow fluid layer forced electromagnetically \citep{Cieslik2010}. Downward motions are concentrated within thin, elongated structures coinciding with strain-dominated (hyperbolic) regions of the horizontal flow. Particles tend to spend relatively short times in such regions, since they are unstable from a dynamical systems perspective \cite{Kadoch2011}. In contrast, upward motions are located within the vortices, i.e. inside  the vorticity-dominated (elliptical) regions of the horizontal flow. Particles released in those regions stay there for long times due to the trapping properties of the elliptical points \cite{Babiano1993,Kadoch2011}, resulting in a delay in the homogenization of the particles within the carrier flow \cite{Provenzale1999}. Consequently, the particles keep their initial upward (positive) direction for longer times. This explains the ability of the updrafts to retain particles for longer times compared to the downdrafts. Hence, the asymmetry in the characteristics of the upward and downward flows directly determines the amount of material transported to the top and to the bottom of the layer, the rate at which this transport occurs, and the horizontal homogenization rate.

The asymmetry in the vertical dispersion, as well as in the vertical flows, is affected by the fluid layer depth, characterized by the aspect ratio $\delta$, and the flow strength, characterized by the Reynolds number $Re_F$. In particular, the degree of transport asymmetry responds to the parameter $Re_F\delta^3$ (provided that the flow is within the inertial regime and time-dependent). In flows with small $Re_F\delta^3$, the vortices (and hence updrafts) become even larger, while the structures where strain is dominant (and hence downdrafts) become thinner and more elongated. In these circumstances, the above-mentioned differences between the upwards and downwards transport are further enhanced (see Fig.~\ref{fig:particle_traj} for a visual cue). As $Re_F\delta^3$ increases, the flow becomes more unsteady. This leads to more vortex-vortex interactions, which result in the modification of flow patterns by, for example, deformation of  the vortex shape or merging of two neighboring vortices. Concurrently, smaller flow structures emerge, which are minimally correlated with either the updrafts or the downdrafts. Consequently, there is a reduction in the correlation between updrafts and vortices, as well as between downdrafts and strain-dominated structures, which explains the reduction in transport asymmetry. Under highly unsteady conditions (larger $Re_F\delta^3$), the flow becomes disorganized and disintegrate into smaller, 3D structures. At this point, the updrafts and downdrafts take the form of elongated filaments of similar strength and size (see Fig.~\ref{fig:wfields_deltas}). These filaments exhibit minimal correlation with a specific region of the horizontal flow. As result, the asymmetry is eliminated as particles are transported by updrafts and downdrafts concentrated within these filaments.

The findings described in the present paper are at least qualitatively relevant for  environmental shallow flows because our simulated flows exhibit basic characteristics of their environmental counterparts. The most relevant, apart from shallowness, is the presence of large-scale vortices (although laminar). Therefore, the present results can reveal generic mechanisms of vertical transport within shallow vortices that might also be active in the environmental context. Thus, natural vortices affect vertical dispersion by firstly advecting large amounts of material upwards, and secondly, trapping this material for prolonged periods.  These effects decrease if the vortex is perturbed by the overall flow conditions, thereby modifying the properties of vertical transport. However, particles within environmental flows are also redistributed and spread by small-scale turbulence. Typically, to model the presence of turbulence in the particles motion, a random velocity is added to the velocity of each particle \cite{ Griffa1996}. This approach was not implemented in our particle tracking. Nevertheless, this presents an opportunity to investigate the influence of the small-scale turbulent motions on the vertical transport asymmetry, which is essential to fully explain this dispersive process.

Finally, it should be noted that only passive particles that follow the flow perfectly were employed in this study. However, plastics, sediments or microorganisms float, sink or swim and can be affected by inertia and their finite-size causing their motion to deviate from that of passive particles. Hence, for environmental applications, it is relevant to understand how these additional properties interact with the vertical flows described in the present study and affect the vertical transport.

\section*{Acknowledgements}
L.M.F.R. gratefully acknowledges financial support from the Consejo Nacional de Humanidades, Ciencias y Tecnolog\'ias (CONAHCYT, M\'exico) through a scholarship grant (No. 710021).

\appendix
\section{Dimensional analysis for obtaining the viscous and inertial  flow regimes \label{appendix:dimensional}}

\subsection{Viscous regime}

For a weak forcing and correspondingly small Reynolds numbers, inertia can be neglected, and the dominant balance is between the external body force and the viscous force, i.e.,

\begin{equation}
\nu\nabla^2\mathbf{u}\approx\dfrac{\mathbf{f}}{\rho}. \label{eq:viscous_body_balance}
\end{equation}
Due to small depth, viscous dissipation due to bottom friction becomes the dominant  dissipation mechanism. In this scenario, the vertical profile of the horizontal flow field can be approximated with a Poiseuille-like profile, such as $u(x,y,z)=u^*(x,y)\sin(\pi z/H)$ (and a similar definition for $v$). Consequently, the order of magnitude of the viscous force is given by

\begin{equation}
\nu\nabla^2\mathbf{u}\approx\nu\dfrac{\partial^2\mathbf{u}}{\partial z^2}\sim \dfrac{\pi^2\nu\mathcal{U}}{H^2},
\end{equation}
considering that $\partial/\partial z\gg \partial/\partial x,\,\partial/\partial y$. On the other hand, the magnitude of the body force is of order $\mathcal{F}/\rho$. Therefore, the balance in Eq.~\eqref{eq:viscous_body_balance} becomes

\begin{equation}
    \dfrac{\pi^2\nu\mathcal{U}}{H^2}\sim \dfrac{\mathcal{F}}{\rho},
\end{equation}
which is equivalent to 

\begin{equation}
    Re\sim \dfrac{Re_F^2 \delta^2}{\pi^2}. \label{eq:Re_viscous_reg}
\end{equation}

\subsection{Inertial regime}

For relatively strong forcing and correspondingly large Reynolds numbers, it is assumed that the external body force and the inertia forces are of the same order, so

\begin{equation}
    (\mathbf{u}\cdot\boldsymbol{\nabla})\mathbf{u}\approx\dfrac{\mathbf{f}}{\rho} \label{eq:inertia_body_balance}
\end{equation}
with the order of the inertial term given by
\begin{equation}
    (\mathbf{u}\cdot\boldsymbol{\nabla})\mathbf{u}\sim \dfrac{\mathcal{U}^2}{L_f}.
\end{equation}
Therefore, the balance in Eq.~\eqref{eq:inertia_body_balance} implies that

\begin{equation}
    \dfrac{\mathcal{U}^2}{L_f}\sim \dfrac{\mathcal{F}}{\rho},
\end{equation}
which is equivalent to

\begin{equation}
        Re\sim Re_F. \label{eq:Re_inertial_reg}
\end{equation}

Besides, there should be a balance between the external body force and the viscous force, because the energy input must be balanced by the viscous dissipation to obtain a statistically steady flow. However, the body forces cannot be of the same order of the viscous forces if the flow has a Poiseuille-like vertical profile. Hence, the velocity is assumed to vary with a vertical length scale $h$ such that 
\begin{equation}
\nu\nabla^2\mathbf{u}\approx\nu\dfrac{\partial^2\mathbf{u}}{\partial z^2}\sim \dfrac{\nu\mathcal{U}}{h^2}.
\end{equation}
To obtain $h$, we use the balance of the inertia and viscous forces, resulting in

\begin{equation}
    h\sim \dfrac{H}{Re^{1/2}\delta}.
\end{equation}
The scale $h$ can be interpreted as a non-dimensional thickness of the boundary layer at the no-slip bottom.



\bibliographystyle{elsarticle-num-names} 
\bibliography{references}





\end{document}